\documentclass[12pt]{article}
\usepackage{graphicx}
\usepackage{axodraw}
\usepackage[dvips]{color}
\usepackage{epsfig}

\def\r2{\sqrt 2}
\def\beq{\begin{equation}}
\def\eeq{\end{equation}}
\def\beqn{\begin{eqnarray}}
\def\eeqn{\end{eqnarray}}

\def\sinW2{\sin^2\theta_W}

\def\mz2{M_{z}^2}
\def\c2b{\cos 2\beta}

\def\m#1{{\tilde m}_#1}

\def\mz{M_z}

\def\sec2w{sec^2\theta_W}
\def\m12{m_{\frac{1}{2}}}
\def\12{$\frac{1}{2}$}
\def\m0{$m_0$}
\def\bsg{b \rightarrow s \gamma}
\def\br{BR($\bsg$)}
\def\mic{{\tt micrOMEGAs }}
\begin{document}
\baselineskip 18pt
\def\today{\ifcase\month\or
 January\or February\or March\or April\or May\or June\or
 July\or August\or September\or October\or November\or December\fi
 \space\number\day, \number\year}
\def\thebibliography#1{\section*{References\markboth
 {References}{References}}\list
 {[\arabic{enumi}]}{\settowidth\labelwidth{[#1]}
 \leftmargin\labelwidth
 \advance\leftmargin\labelsep
 \usecounter{enumi}}
 \def\newblock{\hskip .11em plus .33em minus .07em}
 \sloppy
 \sfcode`\.=1000\relax}
\let\endthebibliography=\endlist

\begin{titlepage}

\begin{center}
{\large {\bf An Improved Analysis of $b\rightarrow s\gamma$ in Supersymmetry}}\\

\vskip 1.5 true cm

\renewcommand{\thefootnote}
{\fnsymbol{footnote}}
Mario E. G\'omez$^{a}$,  Tarek Ibrahim$^{b,c}$, Pran Nath$^{c}$ and 
 Solveig Skadhauge$^d$  
\vskip 0.8 true cm
\noindent
{\it a. Departamento de F\'{\i}sica Aplicada, Facultad de Ciencias 
Experimentales,}\\
{\it Universidad de Huelva, 21071 Huelva, Spain}\\
{\it b. Physics Department, American University of Beirut, Beirut,
Lebanon\footnote{Current address of T.I.}} \\ 
{\it c. Department of Physics, Northeastern University,
Boston, MA 02115-5000, USA.} \\
{\it d. Instituto de F\'{\i}sica,     
Universidade de S\~{a}o Paulo, 05315-970 S\~{a}o Paulo, SP, Brazil.}
\end{center}
      
\vskip 1.0 true cm
\centerline{\bf Abstract}
\medskip

An improved analysis of the $b\rightarrow s+\gamma$ decay 
in the minimal flavor violating case  is given 
taking into account additional contributions in the supersymmetric 
sector which enter in the next-to-leading-order (NLO) and are 
enhanced by $\tan\beta$ factors. Specifically, we compute a set 
of twenty one-loop diagrams to give the most complete analysis 
to date of  the NLO supersymmetric corrections.
These  modifications are computed from the effective charged Higgs  
and neutral Higgs  couplings involving twelve loop diagrams for the 
charged Higgs sector and eight loop diagrams for the neutral Higgs 
sector.  
While the computations of these corrections are available in the literature, 
 their full forms including the complex phase dependence has not be considered.
  Our analysis takes account of the full allowed set of 
twenty one-loop diagrams  and is more general since it 
also includes the full dependence on CP phases in non 
universal sugra and MSSM models. 
A numerical analysis is carried out to estimate the size of the 
corrections to $b\rightarrow s+\gamma$. We also briefly discuss 
the implications of these  results for the search for supersymmetry.

\end{titlepage}

\section{Introduction}

One of the most severe phenomenological constraints on 
supersymmetric (SUSY) models arises from the measurement of the 
inclusive rare decay $B \rightarrow X_s \gamma$. 
This decay only occurs at the one-loop level in the 
standard model(SM)\cite{sm}, 
and therefore the supersymmetric radiative corrections are 
important and might even be of the same order of magnitude 
as the SM contribution (For early work on supersymmetric
contributions to $b\rightarrow s\gamma$ and implications see 
Refs.\cite{history,Baer:1997jq}).
In this paper we carry out an improved analysis of the branching ratio
BR($b \rightarrow s \gamma$) assuming the extended 
minimal-flavor-violation
(EMFV).  By EMFV we mean
that the squark and quark mass matrices 
are diagonalized with the same unitary transformation, in which 
case the only source of flavor violation is the 
Cabibbo-Kobayashi-Maskawa (CKM) matrix 
but CP violation can arise in our model from both the CKM 
matrix and also from the soft susy breaking parameters.  
The strong constraints on flavor changing neutral current, 
indeed suggest a kind of organizing principle like EMFV 
for the case of softly broken supersymmetry. 

The new results presented in this paper consist of the 
complete calculation of the supersymmetric one-loop corrections 
to the Higgs sector couplings that enter into the calculation 
of the next-to-leading-order contributions to 
BR($\bsg$) through corrections to vertex factors. 
These beyond-leading-order SUSY corrections are parameterized 
by three $\epsilon$'s; $\epsilon_b(t)$, $\epsilon_t(s)$ 
and $\epsilon_{bb}$ and can have large effects due to  
contributions that are enhanced by factors of $\tan\beta$. 
In this paper we  derive the $\tan\beta$ enhanced  
as well as the $\tan\beta$ non-enhanced contributions. 
Of course  there exist two-loop (NLO) 
supersymmetric corrections beyond the ones parametrized by the 
$\epsilon$'s. However, such NLO corrections are generally small or can 
be absorbed in a redefinition of the SUSY 
parameters~\cite{Degrassi:2000qf,Carena:2000uj}. 
As is well known the precision theoretical analyses of 
sparticle masses and couplings are strongly affected by 
the $b\rightarrow s\gamma$ constraint and 
such predictions would be tested at colliders in the future. 
The above provides the motivation for an improved 
$b\rightarrow s\gamma$ analysis which is the
purpose of this analysis. 

The current average value for the BR($b\rightarrow s \gamma$)   
from the experimental data~\cite{bsgdata} is,   
\begin{equation}
\rm{BR}(b\rightarrow s \gamma)=
(355 \pm 24_{-10}^{+9}\pm 3)\times 10^{-6} \;,
\label{bsg1}
\end{equation}
by the {\it Heavy Flavor Averaging Group}~\cite{hfag}. 


The standard model result depends sensitively on the QCD 
corrections~\cite{Chetyrkin:1996vx} and we  will use the 
value~\cite{Gambino:2001ew}
\beqn
  BR(B\rightarrow X_s\gamma) = 
(3.73\pm .30)\times 10^{-4} \ ,
\eeqn

which takes into account NLO QCD corrections. 
In this analysis we largely follow the analysis of the \mic
group~\cite{Belanger:2004yn}, in the computation of the \br, 
with exception of the calculation of the beyond-leading order 
SUSY corrections. Further, we extend to the case of non-zero
CP-phases. In the following we give  the essential basics  of the
analysis and refer the reader to the previous literature  for
more details (see, e.g, Ref.~\cite{Belanger:2004yn} and references therein.).
The theoretical analysis of $b\rightarrow s\gamma$ decay is based 
on the following effective Hamiltonian
\beqn 
H_{eff} = 
-\frac{4 G_F}{\sqrt{2}} V^*_{ts} V_{tb} \sum_{i=1}^{8} C_i(Q) O_i(Q) 
\eeqn
where $V_{tb}$ and $V_{ts}$ are elements of the CKM matrix,  
$O_i(Q)$ are the operators defined below and $C_i(Q)$ are the Wilson 
coefficients evaluated at the scale Q.
The only Wilson coefficients that contribute are $C_2$, $C_7$ and $C_8$
and the corresponding operators  are defined as follows 
(see e.g., Ref.\cite{Chetyrkin:1996vx})
\beqn 
O_2 &=&     (\bar{c}_L \gamma^{\mu}     b_L) (\bar{s}_L \gamma_{\mu} c_L)  
\nonumber \\
O_7  &=&  \frac{e}{16 \pi^2} m_b (\bar{s}_L \sigma^{\mu \nu} b_R) F_{\mu \nu} 
\\   
O_8  &=&  \frac{g_s}{16 \pi^2} m_b (\bar{s}_L \sigma^{\mu \nu} T^a
b_R) G_{\mu \nu}^a \nonumber
\eeqn
Here $e$ is the magnitude of the electronic charge,
  $g_s$ is the strong coupling constant,  
$T^a$ (a=1,..,8) are the generators of $SU(3)_C$ 
and  $G^a_{\mu\nu}$ are the gluonic field strengths. 
As is well known the decay width $\Gamma(B\rightarrow X_s\gamma)$ 
has an $m_b^5$ dependence and thus subject to significant uncertainty 
arising from the uncertainty in the b quark mass measurement. 
However, the semileptonic decay width 
$\Gamma(B\rightarrow X_ee\bar \nu)$ also has the same $m_b^5$ dependence
but is experimentally well determined. For this reason one considers the
ratio of the two decay widths where the strong $m_b$ dependence
cancels out. The ratio of interest including the photon detection 
threshold is defined by~\cite{Chetyrkin:1996vx,kagan} 
 \beqn
   R_{\rm th}(\delta) =
   \frac{\Gamma(B\to X_s\gamma)\big|_{E_\gamma>(1-\delta)
         E_\gamma^{\rm max}}}{\Gamma(B\to X_c\,e\,\bar\nu)}
   = \frac{6\alpha}{\pi f(z)}\,\left| \frac{V_{ts}^* V_{tb}}{V_{cb}}
    \right|^2 K_{\rm NLO}(\delta) \,,
\label{rdelta}
\eeqn
where $f(z)=1-8z^2+8z^6-z^8-24z^4\ln z$ is a phase space factor and 
$z=(m_c/m_b)$ is given in terms of pole masses. 
We take $\delta$, which is related to the photon detection threshold, 
to be 0.9 and $\Gamma(B\to X_c\,e\,\bar\nu)$ to be 0.1045.  
$K_{\rm NLO}$ depend on the Wilson coefficients
and is given in the form~\cite{kagan,kagan2} 
\beqn
   K_{\rm NLO}(\delta) &=& \sum_{ \stackrel{i,j=2,7,8}{i\le j} }
    k_{ij}(\delta,Q_b)\,\mbox{Re}\!\left[ C_i^{(0)}(Q_b)\,
    C_j^{(0)*}(Q_b) \right] + S(\delta) \frac{\alpha_s(Q_b)}{2\pi} 
    \mbox{Re} [C_7^{(1)}(Q_b)C_7^{(0)*}(Q_b)] \nonumber \\
    &+& S(\delta)\frac{\alpha}{\alpha_s(Q_b)}\left( 
        2 \mbox{Re}[C_7^{\rm (em)}(Q_b)C_7^{(0)*}(Q_b)] 
        - k^{\rm (em)}(Q_b) |C_7^{(0)}(Q_b)|^2 \right)
\eeqn
where $k_{ij}, S(\delta)$ are as defined in
Ref.\cite{kagan2}, 
and we use the running charm mass $m_c(m_b)$ as suggested 
in Ref.\cite{Gambino:2001ew}. We take the renormalization scale, 
$Q_b$, to be the b-quark mass.
Above the Wilson coefficients have been expanded in terms of 
leading-order and next-to-leading order as follows\cite{kagan}
\beqn
   C_i(Q_b) = C_i^{(0)}(Q_b) + \frac{\alpha_s(Q_b)}{4\pi}\,
   C_i^{(1)}(Q_b) + \frac{\alpha}{\alpha_s(Q_b)} C_i^{({\rm em})}(Q_b)\,.
\eeqn
The coefficients to leading order at the scale of the b-quark 
mass can be obtained from the Wilson coefficients at the electroweak 
scale $Q_W$ by renormalization group evolution such that
\beqn
   C_2^{(0)}(Q_b) &=& \frac 12 \left( \eta^{-\frac{12}{23}}
    + \eta^\frac{6}{23} \right) \,, \nonumber\\
   C_7^{(0)}(Q_b) &=& \eta^\frac{16}{23}\,C_7^{(0)}(M_W)
    + \frac 83 \left( \eta^\frac{14}{23} - \eta^\frac{16}{23} \right)
    C_8^{(0)}(M_W) + \sum_{i=1}^8\,h_i\,\eta^{a_i} \,, \nonumber\\
   C_8^{(0)}(Q_b) &=& \eta^\frac{14}{23}\, (C_8^{(0)}(M_W) + 
   \frac{313063}{363036}) + \sum_{i=1}^4\,\bar h_i\,\eta^{b_i} \,,
\label{newone}
\eeqn
where $\eta=\alpha_s(M_W)/\alpha_s(Q_b)$ and $h_i$, $\bar h_i$,
$a_i$ and $b_i$ are numerical coefficients and are listed in  
Appendix A\cite{Chetyrkin:1996vx}. 
The next-to-leading order contributions and $k^{\rm em}$ 
 are defined 
as in Ref.\cite{Chetyrkin:1996vx,Belanger:2004yn}. 

The main focus of this paper is the next-to-leading-order 
supersymmetric contributions to the Wilson coefficients $C_{7,8}$ 
at the electroweak scale.  
Here $C_{7,8}$ are sums of the Standard Model contribution arising 
from the exchange of the W and from the exchange of the charged Higgs 
and the charginos, so that
\beqn
C_{7,8}(Q_W )= C_{7,8}^{W}(Q_W ) 
 + C_{7,8}^{H^{\pm}}(Q_W ) + C_{7,8}^{\chi^{\pm}}(Q_W ) \;.
\eeqn 
Additonally the gluino exchange contribution has been computed in 
Ref.~\cite{Everett:2001yy}. However,  contributions to the Wilson coefficients 
arising from gluino and neutralino exchange are negligible 
in the MFV scenario. 
Studies of BR($\bsg$) beyond the MFV scenario, by looking 
at the effects from generational squark mixing, has 
recently been performed in Ref.~\cite{bsgNMFV}. 
In the analysis of the supersymmetric contributions to the 
next-to-leading-order we will take into account the 
CP phase dependence. It is now well known that 
large CP phases can appear in SUSY, string and brane models 
while still allowing for the possibility of electric dipole moments 
of the electron, of the neutron and of the $^{199}$Hg atom consistent 
with experiment\cite{na,incancel,olive,chang}. (For the current
experiment on the EDMs see Refs.\cite{eedm,nedm,atomic}.) 
If phases are large they will have important effects on a number of 
phenomena~\cite{cphiggsmass,Christova:2002sw,Ibrahim:2003ca,ibrahim3,
cpdark,gomez,gomez2,Ibrahim:2003tq, Chattopadhyay:2005qu}.

The outline of the rest of the paper is as follows: In Sec.2 we give 
the effective Lagrangian for the charged Higgs and exhibit how the 
corrections $\epsilon_b'(t)$, $\epsilon_t'(s)$ and $\epsilon_{bb}$, 
which bring in $\tan\beta$ factors, enter in the charged 
Higgs Yukawa couplings. In Sec. 3 we exhibit the dependence on 
$\epsilon_b'(t)$, $\epsilon_t'(s)$ and $\epsilon_{bb}$ 
of the Wilson coefficients $C_{7,8}$. 
In Sec.4 we give a comparison of our work with previous ones.
A numerical analysis is given in Sec.5 and we determine regions of the
parameter space where sizeable differences occur using the full 
formulae derived in this paper relative to the partial results of  
some of the previous works.  
Conclusions are given in Sec.6. In appendix A, the 
parameters $h_i$, $\bar h_i$, $a_i$ and $b_i$ that appear 
in Eq. (\ref{newone}) before are listed. In appendix B
we give an analysis of $\epsilon_b'(t)$ by computing the six 
diagrams in Fig.(\ref{epsfig1}). In appendix C we give 
an analysis of $\epsilon_t'(s)$ by computing the six diagrams 
in Fig.(\ref{epsfig2}). In appendix D we give an analysis of 
$\epsilon_{bb}$ by computing the six diagrams in 
Fig.(\ref{epsfig3}) and the two diagrams of Fig.(\ref{epsfig4}).
 
\section{Effective Lagrangian}\label{lagrang}

To discuss the beyond-leading-order supersymmetric contribution 
it is convenient to look at the effective Lagrangian describing 
the interactions of quarks with the charged Higgs fields 
$H^{\pm}$ and the charged Goldstones $G^{\pm}$.
We use the framework of the minimal supersymmetric standard model 
(MSSM) which contains two isodoublets of Higgs bosons.
Thus for the Higgs sector we have
\beqn
(H_1)= \left(\matrix{H_1^1\cr
 H_1^2}\right),~~
(H_2)= \left(\matrix{H_2^1\cr
             H_2^2}\right)
\label{newa}
\eeqn
The components of $H_1$ and $H_2$ interact with the quarks at 
the tree level through
\beqn
-{\cal {L}}= \epsilon_{ij} h_b\bar b_R  H_1^i Q_L^j -
\epsilon_{ij} h_t\bar t_R  H_2^i Q_L^j +H.c 
\label{newb}
\eeqn
The SUSY QCD and the SUSY Electroweak loop corrections 
produce shifts in these couplings and generate new ones as follows
\beqn
-{\cal {L}}_{\rm eff}= \epsilon_{ij} (h_b+\delta h_b^i)\bar b_R  H_1^i
Q_L^j +\Delta h_b^i \bar b_R H_2^{i*} Q_L^i\nonumber\\
-\epsilon_{ij} (h_t+\delta h_t^i)\bar t_R  H_2^i Q_L^j 
+\Delta h_t^i \bar t_R H_1^{i*} Q_L^i
+H.c 
\label{newc}
\eeqn
where 
the complex conjugate is needed
to get a gauge invariant ${\cal {L}}_{eff}$. 
We note that in the approximation 
\beqn
\delta h_f^1 =\delta h_f^2, \nonumber\\
\Delta h_f^1 =\Delta h_f^2
\label{newd}
\eeqn
one finds that 
Eq.(\ref{newc}) preserves weak isospin. This is the approximation 
that is often used in the literature. However, in general the 
equalities of Eq.(\ref{newd}) will not hold and there will be 
violations of weak isospin. It has been demonstrated that the 
weak isospin violation can be quite significant, i.e, as mush 
as $40-50\%$ or more of the total loop correction to the 
Yukawa coupling \cite{Ibrahim:2003tq}.

The typical supersymmetric loop that contributes to the 
shifts in the couplings $\Delta h_f^i$ and $\delta h_f^i$ contains
one heavy fermion $f$ and two heavy scalars $S_1$ and $S_2$ 
or one heavy scalar $S$ and two heavy fermions $f_1$ and $f_2$.
The basic integral that enters in the first case is
\beq
I_1=\int \frac{d^4k}{(2\pi)^4}
\frac{m_f +\not\!k}{(k^2-m^2_f)(k^2-m^2_{S_1})(k^2-m^2_{S_2})}
\eeq
The basic integral that enters in the second case is
\beq
I_2=\int \frac{d^4k}{(2\pi)^4}
\frac{(m_{f_1} +\not\!k)(m_{f_2} +\not\!k)}
{(k^2-m^2_{f_1})(k^2-m^2_{f_2})(k^2-m^2_{S})}
\eeq
The largest finite parts of these integrals that contribute to the 
vertex corrections, in the zero external momentum analysis would read
\beqn
I_1=\frac{1}{16 \pi^2 m_f} H(\frac{m^2_{S_1}}{m^2_f},
\frac{m^2_{S_2}}{m^2_f})\nonumber\\
I_2=\frac{m_{f_1} m_{f_2}}{16 \pi^2 m^2_S} H(\frac{m^2_{f_1}}{m^2_S},
\frac{m^2_{f_2}}{m^2_S})
\eeqn
where the function $H$ is given by
\beq
H(x,y)=\frac{x}{(1-x)(x-y)}\ln x + \frac{y}{(1-y)(y-x)}\ln y 
\eeq
in case $x\neq y$ and
\beq
H(x,y)=H(x)=\frac{1}{(x-1)^2}[1-x+\ln x]
\eeq
for the case $x=y$.

Electroweak symmetry is broken spontaneously by giving vacuum 
expectation value $v_1/\sqrt{2}$ to $H_1^1$ and 
$v_2/\sqrt{2}$ to $H_2^2$. Then the mass terms for the quarks 
arising from Eq.(\ref{newc}) would be
\beq
-{\cal {L}}_m=m_b \bar{b}_R b_L + m_t \bar{t}_R t_L +H.c
\eeq
with $m_b$ and $m_t$ related to $h_b$ and $h_t$ as follows
\beqn
h_b=\frac{\sqrt{2} m_b}{v_1 (1+\epsilon_{bb} \tan\beta)}\nonumber\\
h_t=\frac{\sqrt{2} m_t}{v_2 (1+\epsilon_{tt} \cot\beta)}
\label{newe}
\eeqn
where 
\beqn
\epsilon_{bb}=\frac{\Delta h_b^2}{h_b}
+\cot\beta \frac{\delta h_b^1}{h_b}\nonumber\\
\epsilon_{tt}=\frac{\Delta h_t^1}{h_t}+\tan\beta \frac{\delta h_t^2}{h_t}
\label{newf}
\eeqn

The electroweak eigenstates of charged Higgs interaction 
with quarks in Eq.(\ref{newc}) is
\beqn
{\cal {L}}_{\rm eff}= (h^*_b+\delta h_b^{2*})\bar t_L b_R H_1^{2*}
-\Delta h_b^{1*} \bar{t}_Lb_R H_2^1\nonumber\\
+(h_t+\delta h_t^{1})\bar t_R b_L H_2^{1}
-\Delta h_t^{2} \bar{t}_R b_L H_1^{2*}
+H.c 
\label{newg}
\eeqn
By going from the electroweak eigenstates basis to the 
mass eigenstate $H^+$ and $G^+$ basis
\beqn
H_1^{2*}=\sin\beta H^+ -\cos\beta G^+\nonumber\\
H_2^{1}=\cos\beta H^+ +\sin\beta G^+
\eeqn
and by using Eq. (\ref{newe}) one gets
\beqn
{\cal {L}}_{\rm eff} &=& \!\! \frac{g}{\sqrt 2 M_W} G^+ 
\{ m_t \frac{1+\epsilon_t(b) \cot\beta}{1+\epsilon_{tt}\cot\beta}\bar t_R b_L 
-m_b \frac{1+\epsilon_b'(t)\tan\beta}{1+\epsilon_{bb}^*\tan\beta}
\bar t_L b_R\} \\
&+&\frac{g}{\sqrt 2 M_W} H^+ \{  m_t \frac{1+\epsilon^{\prime}_t(b) 
\tan\beta}{1+\epsilon_{tt}\cot\beta} \cot\beta \bar t_R b_L 
+ m_b \frac{1+\epsilon_b(t)\cot\beta}{1+\epsilon_{bb}^*\tan\beta}
\tan\beta \bar t_L b_R \} + H.c. \nonumber 
\label{newh}
\eeqn
with
\beqn
\epsilon_t(b) &=& 
\frac{\Delta h_t^2}{h_t}+\tan\beta \frac{\delta h_t^1}{h_t}\nonumber\\
\epsilon^{\prime}_b(t) &=& \frac{\Delta h_b^{1*}}{h^{*}_b}
+\cot\beta \frac{\delta h_b^{2*}}{h^{*}_b}\nonumber\\
\epsilon^{\prime}_t(b) &=& -\frac{\Delta h_t^2}{h_t}
+\cot\beta \frac{\delta h_t^1}{h_t}\nonumber\\
\epsilon_b(t) &=& -\frac{\Delta h_b^{1*}}{h^{*}_b}
+\tan\beta \frac{\delta h_b^{2*}}{h^{*}_b}
\label{newi}
\eeqn

For flavor mixing 
to be considered we should have worked out the analysis 
for three generations of quarks 
from the beginning. Thus the general effective largrangian would read
\beqn
{\cal {L}}_{\rm eff} &=& \frac{g}{\sqrt 2 M_W} G^+ \{\sum_d m_t V_{td}  
\frac{1+\epsilon_t(d) \cot\beta}{1+\epsilon_{tt}\cot\beta}\bar t_R d_L 
-\sum_u m_b V_{ub} \frac{1+\epsilon_b^{\prime}(u)
\tan\beta}{1+\epsilon_{bb}^*\tan\beta}
\bar u_L b_R\} \nonumber \\
&+&\frac{g}
{\sqrt 2 M_W} H^+ \{\sum_d m_t V_{td}\frac{1+\epsilon^{\prime}_t(d)\tan\beta}
{1+\epsilon_{tt}\cot\beta} \cot\beta \bar t_R d_L \\
&+&\sum_u m_b V_{ub} \frac{1+\epsilon_b(u)\cot\beta}
{1+\epsilon_{bb}^*\tan\beta} \tan\beta \bar u_L b_R \}
+ H.c.  \nonumber
\label{newj}
\eeqn
where $V_{qq'}$ here are the radiatively corrected CKM
matrix elements.
\footnote{For a precise analysis of introducing CKM matrix elements
into the Lagrangian and their radiative corrections, see \cite{babunew}.}

The terms with $\epsilon_{tt}$ can be ignored since the 
radiative corrections for the top quark mass are typically 
less than $1\%$ \cite{Ibrahim:2003ca}. 
 As in \cite{Degrassi:2000qf,Belanger:2004yn,Demir:2001yz} 
we will ignore the terms with $\epsilon_t(d)$ and $\epsilon_b(u)$. 
This is a good approximation in the large $\tan\beta$ region, 
and for small values of $\tan\beta$ these $\epsilon$'s have little or
no influence on the rate of $\bsg$.

\section{Wilson Coefficients}

Using the above Lagrangian for the interactions of quarks 
with the charged Higgs and the charged Goldstone bosons 
along with the Lagrangian that describes the interaction 
of quarks with $W^{\pm}$ bosons:
\beq
{\cal {L}}= g \sum_d V_{td} \bar t_L \gamma^{\mu} d_L W^+_{\mu}
+ H.c,
\eeq
 the 
contributions to $C_{7,8}$ from the $W$-boson and 
from the charged Higgs are given by: 
\beqn
C_{7,8}^{W}(Q_W ) &=& F_{7,8}^{(1)}(x_t) + 
\frac{(\epsilon_{bb}^*-\epsilon_{b}'(t)) \tan\beta}
{1+\epsilon_{bb}^*\tan\beta} F_{7,8}^{(2)} (x_t)
\\
C_{7,8}^{H^{\pm}}(Q_W ) &=& \frac{1}{3\tan^2\beta} F_{7,8}^{(1)}(y_t)
+  
\frac{1+\epsilon_{t}'(s)^*\tan\beta}{1+\epsilon_{bb}^*\tan\beta} 
 F_{7,8}^{(2)}(y_t) 
\eeqn
where $x_t$ and $y_t$ are defined  by 
\beqn
x_t= \frac{m_t^2(Q_W)}{M_W^2}, ~~y_t= \frac{m_t^2(Q_W)}{M_H^2}
\eeqn
and $F_{7,8}^{(1)}$ and  $F_{7,8}^{(2)}$ are given by 
\beqn
F_7^{(1)}(x) &=& \frac{x(7-5x-8x^2)}{24(x-1)^3}
+\frac{x^2(3x-2)}{4(x-1)^4}\ln x  \nonumber\\ 
F_7^{(2)}(x) &=& \frac{x(3-5x)}{12(x-1)^3}
+\frac{x(3x-2)}{6(x-1)^3}\ln x \nonumber\\
F_8^{(1)}(x) &=& \frac{x(2+5x-x^2)}{8(x-1)^3}
-\frac{3x^2}{4(x-1)^4}\ln x  \nonumber\\
F_8^{(2)}(x) &=& \frac{x(3-x)}{4(x-1)^3} -\frac{x}{2(x-1)^3}\ln x
\eeqn
In the limit where all the supersymmetric particles becomes heavy, 
the SUSY correction to the $W$ contribution vanishes. Thus, in this 
decoupling limit one finds $\epsilon_{bb}^*=\epsilon_b^{\prime}(t)$. 

The chargino exchange contribution to $C_{7,8}$ with 
the beyond-leading-order SUSY corrections, has been derived in 
Ref.\cite{Degrassi:2000qf} and extended to the case of non-zero 
CP-phases in Ref.\cite{Demir:2001yz}. We have 
\beqn
C_{7,8}^{\chi^{\pm}}(Q_s ) &=& 
- \sum_{k=1}^2 \sum_{i=1}^2 
\{ \frac{2}{3} |r_{ki}|^2 \frac{M_W^2}{m_{\tilde t_k}^2}
F_{7,8}^{(1)}(\frac{m_{\tilde t_k}^2}{m_{\chi_i^{\pm}}^2}) 
+ r_{ki}^*r_{ki}' \frac{M_W}{m_{\chi_i^{\pm}}} 
F_{7,8}^{(3)}(\frac{m_{\tilde t_k}^2}{m_{\chi_i^{\pm}}^2})\}
\nonumber\\
&& + \sum_{i=1}^2 
\{ \frac{2}{3} |\tilde r_{1i}|^2\frac{M_W^2}{m_{12}^2} F_{7,8}^{(1)}
(\frac{m_{12}^2}{m_{\chi_i^{\pm}}^2 })
+ \tilde r_{1i}^*\tilde r_{1i}' \frac{M_W}{m_{\chi_i^{\pm}}} 
F_{7,8}^{(3)}(\frac{m_{12}^2}{m_{\chi_i^{\pm}}^2 })\}
\label{chargino}
\eeqn
where $Q_s$ is the soft SUSY scale and $m_{12}$ is the mass of 
the first and second generation up-type squarks, 
which we take to be identical. Further, 
\begin{equation}
r_{ij}= D_{t1i}^{*}V^{*}_{j1} 
-\frac{m_t(Q_s)}{\sqrt 2 M_W\sin\beta} 
D_{t2i}^{*} V^*_{j2} \;\;,\;\;\;
r_{ij}'= \frac{ D_{t1i}^{*} U_{j2}} {\sqrt 2 \cos\beta 
(1+\epsilon_{bb}^*\tan\beta)}
\label{rij}
\end{equation}
and where 
$\tilde r_{ij}$ and $\tilde r_{ij}'$  are obtained  from $r_{ij}$
and $r_{ij}'$ by setting the matrix $D_t$ to unity. Finally the loop functions 
$F_{7,8}^{(3)}(x)$ appearing in Eq.(\ref{chargino}) are given by 
\begin{equation}
F_7^{(3)}(x) = \frac{(5-7x)}{6(x-1)^2} +\frac{x (3x-2)}{3(x-1)^3}\ln x 
\;\;,\;\;\;\;\;
F_8^{(3)}(x) = \frac{(1+x)}{2(x-1)^2} -\frac{x}{(x-1)^3}\ln x
\end{equation}
The value of the chargino contribution at the scale $Q_W$ is computed  
as in Ref.\cite{Degrassi:2000qf}, where we use $\beta_0=-7$
corresponding to six flavors. 
Only the chargino contribution may give a CP-violating contribution 
at the leading order. However, as the $\epsilon$'s may be complex 
all three contributions; the W, the charged Higgs as well as 
the chargino, may be complex at NLO order. 
We note that all the NLO SUSY corrections scales with 
$1/(1+\epsilon_{bb}^*\tan\beta)$. 
To complete the analysis what remains to be done is the computation of 
$\epsilon_b'(t)$, $\epsilon_t'(s)$ and $\epsilon_{bb}$ and as 
mentioned above we will compute these in the zero external 
momentum analysis~\cite{carena2002}. However, we will calculate all 
one-loop SUSY QCD and SUSY electroweak corrections to these 
for any $\tan\beta$. 
We collect the expressions for these corrections in Appendices B, C and D. 
  While analyses for these  exist in the literature they are not fully general 
when CP phases are present and the soft parameters are in general complex.

\section{Comparison with previous works}

In this section we compare our results with previous works
\footnote{A brief comparison of partial 
analysis of the $\epsilon$'s with  previously works was 
given in Ref.\cite{gomez2}}. We start by comparing our 
results with the work  of \cite{BCRS} (BCRS), as this 
analysis is the most complete  of the previous works.  
It includes the exact one-loop results for the $\epsilon$'s, 
but only in the limit of CP conservation.
Thus BCRS considered the one-loop corrections to the vertex 
of the charged Higgs and charged Goldstones with quarks 
in their Figure 4. 
The  $(\Delta F_L^k)^{JI}$ and $(\Delta F_R^k)^{JI}$ in 
Eqs. (A.8) and (A.9) of Ref.\cite{BCRS},
where $J=1,2,3$ for $u,c,t$ and $I=1,2,3$ for $d,s,b$ are
 related to our 
$\epsilon$'s as follows:
\beqn
(\Delta F_L^1)^{JI}=V_{JI} h_J \sin\beta \epsilon^{\prime}_J(I)\nonumber\\
(\Delta F_L^2)^{JI}=V_{JI} h_J \cos\beta \epsilon_J(I)\nonumber\\
(\Delta F_R^1)^{JI}=V_{JI} h_I^* \cos\beta \epsilon_I(J)\nonumber\\
(\Delta F_R^2)^{JI}=-V_{JI} h_I^* \sin\beta \epsilon^{\prime}_I(J)
\label{epsilons_buras}
\eeqn
The first and third lines of the above set are for the charged Higgs 
$H^+$ and the second and fourth are for the charged Goldstone $G^+$.
The relations relevant for the current analysis are the first and the 
fourth ones and thus we will explicitely check the validity of these.

 In order to compare the $\epsilon$'s of our appendices B 
  and C and vertex corrections presented in the appendix A.3 in 
Ref.\cite{BCRS}, we first establish a dictionary connecting the notation
in the two works.
Thus 
the form factors in the two works 
are related to each other by  $x C_0(x,y,z)= - H(\frac{y}{x},\frac{z}{x})$,
where $H(\frac{y}{x},\frac{z}{x})$ is the form factor used 
in our work and $C_0(x,y,z)$ is the form factor used by BCRS. 
The diagonalizing matrices $Z_{-}^{ij} (Z_{+}^{ij})$ of BCRS 
correspond to our $U^{*}_{ji} (V^{*}_{ji})$ and $Z_{N}^{ij}$ of BCRS 
corresponds to our $X_{ij}$. In our analysis we did not consider 
flavor mixing in the squark sector, so the squark mass-squared 
matrices are $2 \times 2$ and not $6 \times 6$ ones. 
Thus in the case of $\epsilon^{\prime}_b(t)$, the sum in the 
expressions for $\Delta F$ is over the squark mass eigenstates
of the third generation; In another words we sum over the third 
and sixth entries in their matrices. So the $Z_D^{IJ*}$ of BCRS 
corresponds to our $D_{bij}$, $Z_U^{IJ}$ of BCRS corresponds to 
our $D_{tij}$ with $I,J=3i,3j$.
In the case of  $\epsilon^{\prime}_t(s)$ we sum, for the $\tilde{s}$ 
squark over the second and fifth entries in their matrices and the 
$Z_D^{IJ*}$ of BCRS corresponds to our $D_{sij}$ with $I,J=3i-1,3j-1$. 
 As BCRS follows the conventions of Ref.~\cite{Rosiek:1989rs}, 
the Higgs coupling $h_b$ in their Lagrangian is our $-h_b$ and their
$h_t$ is equal to ours. The trilinear coupling $A_t$ in their 
superpotential is our $-h_t A_t$ and their  $A_b$ is our $h_b A_b$ as 
can be seen by comparing the superpotential in Sec. 3 of their 
reference [30] and the superpotential we are using which is the 
same as in Eq. (4.15) of Gunion and Haber \cite{haber}. 
Also the elements $Z_H^{ij}$ are defined in section 4 
of Ref.~\cite{Rosiek:1989rs}.
 
We now give details of the comparison.
\noindent
The first term in $(\Delta F_L^1)^{JI}$ of  Eq.(A.8) of 
Ref.\cite{BCRS} has the following correspondence in our notation 
\beq
(\Delta F_L^1)^{JI}_{\rm 1st ~term} \to
V_{JI} h_J  \sin\beta (\epsilon^{\prime(1)}_J(I)
+\epsilon^{\prime(2)}_J(I))
\eeq
In comparing our results  with theirs one finds  that  we have  an  explicit gluino phase dependence
 $\xi_3$ in our analysis. This gives  us the maximum freedom in the  choice  of the independent set
 of phases  to  carry out the analysis in.

\noindent
The second term in $(\Delta F_L^1)^{JI}$ of Eq. (A.8) of
Ref.\cite{BCRS} has the following correspondence in our notation
\beq
(\Delta F_L^1)^{JI}_{\rm 2nd ~term} \to
V_{JI} h_J\sin\beta (\epsilon^{\prime(3)}_J(I)+\epsilon^{\prime(4)}_J(I))
\eeq
Using $e=g \sin\theta_W$ and $g_1=g \tan\theta_W$, one can prove that 
\beqn
(\alpha_{Jk} D_{J1j} -\gamma_{Jk} D_{J2j})
= -\frac{1}{\sqrt 2} V_{uUN}^{RJjk*}\nonumber\\
(\beta^{*}_{Ik} D^{*}_{I1i} +\alpha_{Ik} D^{*}_{I2i})=
 -\frac{1}{\sqrt 2} V_{dDN}^{LIik},
\eeqn
and we find complete agreement for this term.

\noindent
The third term in $(\Delta F_L^1)^{JI}$ of Eq. (A.8) of
Ref.\cite{BCRS} has the following correspondence in our notation 
\beq
(\Delta F_L^1)^{JI}_{\rm 3rd  ~term} \to 
V_{JI} h_J \sin\beta \epsilon^{\prime(5)}_J(I)
\eeq
One can prove that
\beqn
 h_J D_{I1j} V^{*}_{i2} V_{JI} = V_{uDC}^{RJji*}\nonumber\\
 \frac{g}{\sqrt 2} \sin\beta (-\sqrt 2 X_{3k} U^{*}_{i1}
+X_{2k} U^{*}_{i2} +\tan\theta_W X_{1k} U^{*}_{i2})
= V_{NCH}^{Lki1},
\eeqn
and we see that we again agree with BCRS. 
We notice here that our expression does not have the form 
factor $C_2(x,y,z)$. This form factor comes from the $k^2$ term 
in the integral where a loop with two fermions and one scalar is 
integrated. This part diverges and it is used to renormalize the 
parameters of the theory and could be safely ignored as we shall see when we compare with one of their $\epsilon$'s 
later.

\noindent
The fourth term in $(\Delta F_L^1)^{JI}$ in their Eq. (A.8) 
corresponds to our
\beq
(\Delta F_L^1)^{JI}_{\rm 4th  ~term} \to
V_{JI} h_J \sin\beta \epsilon^{\prime(6)}_J(I)
\eeq
One can prove that
\beq
g(V^{*}_{i1} D^{*}_{J1j} - K_J V^{*}_{i2} D^{*}_{J2j})V_{JI}
= -V_{dUC}^{LIji}
\eeq
and we find once again agreement.

Next we compare with  $(\Delta F_R^2)^{JI}$ in Equation (A.9) of BCRS. 
The first term in $(\Delta F_R^2)^{JI}$ has the following
correspondence to our work
\beq
(\Delta F_R^2)^{JI}_{\rm 1st ~term}\to 
-V_{JI} h_I^* \sin\beta (\epsilon^{'(1)}_I(J)+\epsilon^{'(2)}_I(J))
\eeq
The only difference that appear from the comparison is 
that the first term in BCRS should have an extra factor of $e^{i\xi_3}$.
\noindent
The second term in $(\Delta F_R^2)^{JI}$ of Eq.(A.9) in BCRS has the 
following correspondence to our work 
\beq
(\Delta F_R^2)^{JI}_{\rm 2nd  ~term}\to 
-V_{JI} h_I^* \sin\beta (\epsilon^{\prime(3)}_I(J)+\epsilon^{\prime(4)}_I(J))
\eeq
One can prove that
\beqn
(\alpha^{*}_{Ik} D^{*}_{I1j} -\gamma^{*}_{Ik} D^{*}_{I2j})
= -\frac{1}{\sqrt 2} V_{dDN}^{RIjk}\nonumber\\
(\beta_{Jk} D_{J1i} +\alpha^{*}_{Jk} D_{J2i})
= -\frac{1}{\sqrt 2} V_{uUN}^{LJik*},
\eeqn
 and we find agreement between our result and that of BCRS.

\noindent
The third term in $(\Delta F_R^2)^{JI}$ of their Eq. (A.9) has the 
following correspondence to our work
\beq
(\Delta F_R^2)^{JI}_{\rm 3rd ~term}\to -V_{JI} h_I^* \sin\beta 
\epsilon^{\prime(6)}_I(J)
\eeq
One can prove that
\beqn
g(U_{i1} D_{I1j} -K_I U_{i2} D_{I2j}) V_{JI} = -V_{uDC}^{LJji*}\nonumber\\
\frac{g}{\sqrt 2} \sin\beta (\sqrt 2 X^{*}_{4k} V_{i1}
+X^{*}_{2k} V_{i2} +\tan\theta_W X^{*}_{1k} V_{i2})
= -V_{NCH}^{Rki2},
\eeqn
and we find no difference between our equations and theirs for that term.
\noindent
The fourth term in $(\Delta F_R^2)^{JI}$  of their Eq.  (A.9) has the 
following correspondence to our work
\beq
(\Delta F_R^2)^{JI}_{\rm 4th ~term}\to   
-V_{JI} h_I^* \sin\beta \epsilon^{'(5)}_I(J)
\eeq
One can prove that
\beq
g K_I U_{i2} D^{*}_{J1j} V_{JI}= V_{dUC}^{RIji}
\eeq
and comparing their result with ours we find no difference here either.
  To summarize, for the case with no CP phases we find complete agreement
with the work of BCRS. 
However, for the case of CP-violation we have explicit gluino phase dependence.
We note that the BCRS analysis did not take into account the CP violating
effects in the Higgs sector.
 Specifically, it is now known  that 
in the presence of complex phases in the soft SUSY-breaking 
sector, the three neutral Higgs mass eigenstates are 
mixtures of CP-even and CP-odd fields with production and decay 
properties different from those in the CP conserving scenarios. 
The vertices of these mass eigenstates are affected by this mixing 
moreover this mixing can lead to important effects in susy phenomena 
\cite{Seeit}.  We explain now in further detail exactly where 
\cite{BCRS} misses these effects. 
Thus in \cite{BCRS}, the assignment of the neutral Higgs as CP 
even $h^0$ and $H^0$ and CP odd $A^0$ in Appendix A.3 where 
different $\Delta F$ elements of Sec. 2 are calculated does not 
hold for the case with CP phases. Also the decomposition of the 
electroweak eigenstates in Eq. (3.8) of BCRS does not hold for 
the CP violating case. Eqs. of sections (3.3), (6.1.2), (6.2) and Eq. (6.61) 
should also be modified to take this mixing into account. 
We note, however, that the CP mixing effects in the neutral Higgs  sector do 
not affect the $b\to s\gamma$ analysis in this paper.

Continuing  with the comparison of our work with that of BCRS we find 
that in Sec. (3) of \cite{BCRS}, the authors did not take into 
account  violations of isospin in their analysis as they are 
working in the approximation of Eq.(\ref{newd}) above.  
It is known that the effects of violations of isospin can be 
large, and if such violations were included, then in their 
Eqs.(3.3) and (3.14) the corrections $\Delta_d Y_d$, $\Delta_u Y_d$, 
$\Delta_u Y_u$ and $\Delta_d Y_u$ should have a suffix $i$ where $i$ 
labels the element of an isopsin multiplet. In other words instead 
of the above four corrections one should have eight. Specifically the 
corrections that appear in Eqs. (3.35) and (3.41) are in general 
different from those in Eqs. (3.9) and (3.16). We also note 
that Eq. (3.37) of Ref.\cite{BCRS} is derived based on the assumption 
of isospin invariant loop corrections.

However, compared  to the analysis of the works of 
\cite{Degrassi:2000qf,Belanger:2004yn,Demir:2001yz}
the analysis of Ref.\cite{BCRS} is more general as it takes 
into account more loops.
Thus the analysis of Ref.\cite{BCRS} specifically considered  the case where 
two heavy fermions and one heavy scalar are running in the loops.  
So in their Fig.(9)  they considered the additional important 
corrections to the charged Higgs boson couplings to quarks. 
The loop in Fig. (9a) of \cite{BCRS}  corresponds to our loop 2(vi) and 
the loop in Fig. (9b) corresponds to our loop 2(v).   By looking at their expression for 
$\delta_a \epsilon'_{t}(I)$ we note that there is no summation over 
the squark states. Thus the first term of $\delta_a \epsilon'_{t}(I)$ 
corresponds to the case of squark $j=1$ of our $-\epsilon'^{(6)}_{t}(I)$.
Using the following property of the form factor function $H(x,y)$:
\beqn
\frac{m_1 m_2}{m_3^2} H(\frac{m_2^2}{m_3^2},\frac{m_1^2}{m_3^2})=\frac{m_1}{m_2}H(\frac{m_1^2}{m_2^2},\frac{m_3^2}{m_2^2})
\eeqn
one can make the comparison between our expression and theirs.
So in the limit of vanishing Left-Right squark mixings, our expressions limits to
that of \cite{BCRS}. However there is a minor correction to their 
equation even in that limit. Thus in the first line of their expression for $\delta_a \epsilon'_{t}(I)$, $Z_N^{1j}$ should read $Z_N^{4j}$ and their parameter $a^{lj}$ should be modified a little to be:
\beqn
a^{lj}=Z_{-}^{2l}[s_W Z_N^{1j} +c_W Z_N^{2j}]-\sqrt{2} Z_-^{1l} Z_N^{3j} c_W
\label{news}
\eeqn
where the $\cos\theta_W$ factor in their last term is missing.
The second term of $\delta_a \epsilon'_{t}(I)$ correspond to our
$-\epsilon'^{(6)}_{t}(I)$ with $j=2$ squark. By noting that 
\beq
\gamma_{tk}=\frac{2}{3} g X_{1k} \tan\theta_W
\eeq
 our expression reproduces their second term with the minor change  that
$Z_N^{4j}$ should read $Z_N^{1j}$ and with the above new form of the parameter $a^{lj}$ of Eq.(\ref{news}).
One can repeat the same analysis to 
 compare   $\delta_b \epsilon'_{t}(I)$ of \cite{BCRS}  
 with our $-\epsilon'^{(5)}_{t}(I)$. Here also the analysis of Ref.\cite{BCRS}  
 ignores squark mixing and their expression for $\delta_b \epsilon'_{t}(I)$ corresponds to the squark of $j=1$ 
 in our equation. We need the fact that
\beq
\beta^*_{Ik}=\frac{g}{c_W}(\frac{1}{6} s_W X_{1k}
-\frac{1}{2} c_W X_{2k})
\eeq
to reproduce their expression with the minor change of the definition in Eq.(\ref{news}) above.
We note here that the authors did not consider the part of the loop which has the form factor $C_2(x,y,z)$ as mentioned earlier.
Finally, Eq. (5.6) of Ref.\cite{BCRS} calculates $\epsilon'_{t}(I)$
that corresponds to our $-(\epsilon'^{(1)}_{t}(I)+ \epsilon'^{(3)}_{t}(I))$. 
Our expressions are more general and they  limit to Eq.(5.6) of Ref.\cite{BCRS}  if we ignore in our formula,
 the first, third, fourth and fifth terms of $\epsilon'^{(1)}_{t}(I)$ and by ignoring 19 terms in  
 our $\epsilon'^{(3)}_{t}(I)$ as well. 
We should notice here that they are using a different defintion of $A_b$. In this part of the
paper they use $A_b$ to be our $-A_b$ as this could be seen from their footnote 3 of section 5.2.
Also we note that their 
expression for $\epsilon'_{t}(I)$ is only valid for the CP 
conserving scenario. Thus for nonzero CP phases, $\mu$ in the 
first term should read $\mu^*$ and $A_b$ in the second term should 
read $A_b^*$.
However, the assignment of the $Z_N$ matrix elements here is exactly 
like ours and is different from that in the DGG paper 
(See our note after our Eq.(\ref{new33})).\\
Finally we should mention here that, apart from the small differences
mentioned above with the approximate formulae of BCRS, our formulae should
be rather considered as extensions not corrections of them.
 Next we compare our analysis  to  other earlier works where an effective Lagrangian
similar to ours has been used.

\subsection{$\epsilon_b'(t)$}
First we compare our analysis with the work of Demir and Olive 
(DO)~\cite{Demir:2001yz}. We note that the $\epsilon_{tb}$ 
of DO is identical to our $\epsilon_b'(t)$. DO 
computed two one-loop contributions to $\epsilon_b'(t)$, 
which correspond to the contributions $\epsilon_b^{\prime(1)}(t)$ 
and $\epsilon_b^{\prime(3)}(t)$, in the limit of small squark mixings
and large $\tan\beta$. 
Our $\epsilon_b^{\prime(1)}(t)$ in this limit becomes
\beqn
\epsilon_b^{\prime(1)}(t)=
-\sum_{i=1}^2\sum_{j=1}^2 \frac{2\alpha_s}{3\pi} e^{i\xi_3}
|D_{b2j}|^2 |D_{t1i}|^2
\frac{\mu}{|m_{\tilde g}|} H(\frac{m_{\tilde t_i}^2}{|m_{\tilde g}|^2},
\frac{m_{\tilde b_j}^2}{|m_{\tilde g}|^2})
\label{27}
\eeqn
which is the same as the first part of $\epsilon_{tb}$ in 
Ref.\cite{Demir:2001yz}. (We note that $C_q$ of DO 
is our $D_q$ and there is a typo in Ref.\cite{Demir:2001yz}
in that their $|C_{\tilde t}^{2l}|^2$ should be $|C_{\tilde b}^{2l}|^2$).
Next using
\beqn
\frac{2m_t}{m_b}\cot\beta \alpha_{bk}^*\alpha_{tk}^* =h_t^2 X_{3k}^* X_{4k}^*
\eeqn
we find that our $\epsilon_b^{\prime(3)}(t)$ in the limit assumed by
DO takes the form
\beqn
\epsilon_b^{\prime(3)}(t) \simeq \frac{h_t^2}{16\pi^2} \frac{A_t}{m_{\chi_k^0}}
|D_{b1j}|^2 |D_{t2i}|^2
 X_{3k}^* X_{4k}^* H(\frac{m_{\tilde t_i}^2}{m_{\chi_k^0}^2},
\frac{m_{\tilde b_j}^2}{m_{\chi_k^0}^2})
\label{28}
\eeqn
To compare with Ref.\cite{Demir:2001yz} we define
$\alpha_t=h_t^2/4\pi$ and set $C_o=X$. One finds then that the 
overall sign of this term in DO is opposite to ours. 
As will be discussed later the overall sign of this term 
as computed in \mic\cite{Belanger:2004yn} is also in 
disagreement with the sign given by DO, but in agreement 
with our sign as given above. Furthermore, 
in Ref.\cite{Demir:2001yz} $(C_0)_{4i}(C_0^{\dagger})_{3i}$ 
should be $(C_0^*)_{4i}(C_0^*)_{3i}$.  Aside from these corrections, 
the results of Ref.\cite{Demir:2001yz} for $\epsilon'_{b}(t)$ for 
the parts computed, are in agreement with our result.

Next we compare our results with the work of 
Degrassi {\it et. al.} (DGG)~\cite{Degrassi:2000qf}.
Eq.(15) of DGG can be obtained from Eqs.(\ref{27},\ref{28}) 
of our analysis. 
To compare with the results of DGG we have to keep in mind 
that in the analysis of DGG, $A_b$ and $m_g$ are real. 
The relation between our $X$ and the $N$ of DGG is $X^T=N^*$ 
and thus we note that their $N_{4a}N_{a3}^*$ should read 
$N_{a4}N_{3a}$. Moreover, one finds that the overall 
sign of the Yukawa contribution in DGG should be reversed to
agree with our sign.

The analysis of the \mic group~\cite{Belanger:2004yn} takes into account
all the six $\epsilon_b^{\prime}(t)$ contributions but restricted 
to the case of real parameters and using certain approximations. 
The value of $\epsilon_b^{\prime(1)}(t)$ and 
$\epsilon_b^{\prime(3)}(t)$ is identical to the ones of DGG. 
The simplified formulae implemented in \mic 
for $\epsilon^{\prime (2),(4)}_b(t)$ was derived 
in Ref.\cite{Pierce:1996zz} and $\epsilon^{\prime(5),(6)}_b(t)$ 
was derived in Ref.\cite{Carena:1999py}. 

We begin by displaying our $\epsilon_b^{\prime(2)}(t)$ and 
$\epsilon_b^{\prime(4)}(t)$ in the limit of small squark mixings
\beqn
\epsilon_b^{\prime(2)}(t) 
= \sum_{i=1}^2\sum_{j=1}^2 \frac{2\alpha_s}{3\pi} e^{i\xi_3}
|D_{b2j}|^2 |D_{t1i}|^2  \frac{A_b^*}{\tan\beta}
\frac{1}{|m_{\tilde g}|} H(\frac{m_{\tilde t_i}^2}{|m_{\tilde g}|^2},
\frac{m_{\tilde b_j}^2}{|m_{\tilde g}|^2})
\eeqn
\beqn
\epsilon_b^{\prime(4)}(t)=-\sum_{k=1}^4
 \sum_{i=1}^2\sum_{j=1}^2
 \frac{\mu^*}{\tan\beta}
  |D_{b1j}|^2  |D_{t2i}|^2
  X_{3k}^*X_{4k}^*
\frac{h_t^2}{16\pi^2}
\frac{1}{m_{\chi_k^0}} H(\frac{m_{\tilde t_i}^2}{m_{ \chi_k^0}^2},
\frac{m_{\tilde b_j}^2}{m_{\chi_k^0}^2})
\eeqn
To compare with the analysis of \mic we keep in 
mind that in their work $m_g, \mu$ and $ A $ are all real. 
With this restriction our analysis is in agreement 
with Eq.(B.67) of  Ref.\cite{Belanger:2004yn} specifically 
with the sign. (However, $N_{a4}^*$ in Eq.(B.67) should 
read $N_{a4}$). Thus we support their disagreement with 
the work of DO and of DGG as stated after Eq.(B.67) in 
Ref~\cite{Belanger:2004yn}.

Appropriately extended to the complex case the simplified 
formulae for $\epsilon^{\prime (5)}_b(t)$ read 
\begin{equation}
 \epsilon^{\prime (5)}_b(t)=\frac{\alpha(M_{\rm SUSY})}{8s_W^2\pi}\mu M_2 
 \left( \frac{|D_{t11}|^2}{m_{\tilde{t}_1}^2} 
 H(\frac{|M_2|^2}{m_{\tilde{t}_1}^2},\frac{|\mu|^2}{m_{\tilde{t}_1}^2}) +
      \frac{|D_{t12}|^2}{m_{\tilde{t}_2}^2}
 H(\frac{|M_2|^2}{m_{\tilde{t}_2}^2},\frac{|\mu|^2}{m_{\tilde{t}_2}^2})\right) 
\label{epsbt5approx}
\end{equation}
and $\epsilon^{\prime (6)}_b(t)$ read 
\begin{equation}
 \epsilon^{\prime (6)}_b(t)=\frac{\alpha(M_{\rm SUSY})}{4s_W^2\pi}\mu M_2 
 \left( \frac{|D_{b11}|^2}{m_{\tilde{b}_1}^2} 
 H(\frac{|M_2|^2}{m_{\tilde{b}_1}^2},\frac{|\mu|^2}{m_{\tilde{b}_1}^2}) +
       \frac{|D_{b12}|^2}{m_{\tilde{b}_2}^2}
 H(\frac{|M_2|^2}{m_{\tilde{b}_2}^2},\frac{|\mu|^2}{m_{\tilde{b}_2}^2})\right) 
\label{epsbt6approx}
\end{equation}
These formulae are derived by using the corresponding 
formulae for $\epsilon_{bb}$ and the decoupling limit.  
Moreover, one approximates the chargino masses 
by $\mu$ and $M_2$ and neglects mixing matrixes 
and $U(1)$ contributions. 
We have checked numerically that they approximate the 
full formulae given in Eqs.(\ref{epsb5},\ref{epsb6}) of Appendix B 
rather well over most of the complex parameter space. 

\subsection{$\epsilon_{bb}$}
In this section we carry out a similar analysis with the three 
works~\cite{Demir:2001yz,Degrassi:2000qf,Belanger:2004yn} for the 
case of  $\epsilon_{bb}$.
Comparing with the computation of DO we find that 
the QCD part given in Eq.(7) of Ref.\cite{Demir:2001yz} 
is the same as ours in the limit they are considering. 
To compare with the Yukawa part contribution we note 
that their $C_L$, $C_R$ are related to our $U$ and 
$V$ as follows: $C_L^{\dagger}=V$, and $C_R^{\dagger}=U^*$. 
Then using
\beqn
g^2 \frac{m_t}{m_b} \cot\beta K_t K_b =h_t^2
\eeqn
we find agreement with their analysis provide
their $(C_R^{\dagger})_{2j}$ is substituted by  $(C_R^{\dagger})_{j2}$.
Next, comparing with the work of DGG, we agree with 
the QCD part of their Eq.(10) after taking 
account of the fact that they have no CP phases.
To compare the contribution of the chargino in their work 
with ours we note that their $U$ is our $U^*$. Also, 
$V_{a2}$ in their work should read $V_{a2}^*$.
We note that there is also a disagreement between 
DGG and DO on this point taking into account that $C^{\dagger}_L$ in
Ref.\cite{Demir:2001yz} is $V$ in Ref.\cite{Degrassi:2000qf}
and $C^{\dagger}_R$ in  Ref.\cite{Demir:2001yz} corresponds 
to the matrix $U$ of  Ref.\cite{Degrassi:2000qf}.
Finally we compare with the analysis of \mic 
as given in Eq.(B.66) in Ref.\cite{Belanger:2004yn}. 
We agree with their result except that their $V_{a2}$ 
should read $V_{a2}^*$. 
The simplified formulae for $\epsilon^{(1)}_{bb}+\epsilon^{(2)}_{bb}+$
and $\epsilon^{(3)}_{bb}+\epsilon^{(4)}_{bb}$ extended to 
the complex case reads
\begin{equation}
  \epsilon^{(1)}_{bb}+\epsilon^{(2)}_{bb}=\frac{2\alpha_s(M_{\rm SUSY})}{3\pi}
   \frac{(A_b/\tan\beta - \mu^*)}{m_{\tilde{g}}} 
  H(\frac{m_{\tilde{b}_1}^2}{|m_{\tilde{g}}|^2},
    \frac{m_{\tilde{b}_2}^2}{|m_{\tilde{g}}|^2})
\label{bb12}
\end{equation}
and 
\begin{equation}
  \epsilon^{(3)}_{bb}+\epsilon^{(4)}_{bb} =\frac{y_t^2(M_{\rm SUSY})}{16\pi^2} 
  \sum_{a=1,2}U^*_{a2}V^*_{a2}\frac{\mu /\tan\beta-A_t^*}{m_{\chi_a^+}}
  H(\frac{m_{\tilde{t}_1}^2}{m_{\chi_a^+}^2},
    \frac{m_{\tilde{t}_2}^2}{m_{\chi_a^+}^2})
    \label{bb34}
\end{equation}
In \mic the implementation of the terms 7 and 8 are given 
by  $\epsilon^{7}_{bb}=2(\epsilon^{\prime (5)}_b(t))^*$ 
and $\epsilon^{8}_{bb}=(\epsilon^{\prime (6)}_b(t))^*/2$ using the 
results in Eqs.(\ref{epsbt5approx},\ref{epsbt6approx}). 

\subsection{$\epsilon_t'(s)$}
First we compare our analysis with the result of DO, where 
$\epsilon_{ts}$ corresponds to our $\epsilon_t'(s)$.
DO only considered  $\epsilon_t^{\prime (1)}(s)$ and computed this 
in the limits mentioned in the preceding discussion. 
Our result in the same limits is given by
\beqn
\epsilon_t^{\prime(1)}(s)
=\sum_{i=1}^2\sum_{j=1}^2 \frac{2\alpha_s}{3\pi} e^{-i\xi_3}
\mu^* |D_{s1i}|^2 |D_{t2j}|^2
\frac{1}{|m_{\tilde g}|} H(\frac{m_{\tilde s_i}^2}{|m_{\tilde g}|^2},
\frac{m_{\tilde t_j}^2}{|m_{\tilde g}|^2})
\eeqn
Using $D_{s11}\simeq 1$, $D_{s12}\simeq 0$, and $m_{\tilde s_1}^2=Q_{12}^2$,
we get exactly the $\epsilon_{ts}$ of Eq.(7) in Ref.\cite{Demir:2001yz}.
DGG only computed the $\tan\beta$ enhanced QCD and Yukawa 
terms; $\epsilon_t^{\prime(1)}(s)$ and  $\epsilon_t^{\prime(3)}(s)$. 
Our $\epsilon_t^{\prime(1)}(s)+\epsilon_t^{\prime(3)}(s)$ corresponds 
to their Eq.(16) reads
\beqn
\epsilon_t^{\prime(1)}(s) + \epsilon_t^{\prime(3)}(s)
=\sum_{i=1}^2\sum_{j=1}^2 \frac{2\alpha_s}{3\pi} e^{-i\xi_3}
\mu^* |D_{s1i}|^2 |D_{t2j}|^2
\frac{1}{|m_{\tilde g}|} H(\frac{m_{\tilde s_i}^2}{|m_{\tilde g}|^2},
\frac{m_{\tilde t_j}^2}{|m_{\tilde g}|^2})
\nonumber\\
-\frac{h_s^2}{16\pi^2}\sum_{k=1}^4 \frac{A_s^*}{m_{\chi_k^0}} X_{3k}X_{4k}
|D_{s2i}|^2 |D_{t1j}|^2   H(\frac{m_{\tilde s_i}^2}{m_{\chi_k^0}^2},
\frac{m_{\tilde t_j}^2}{m_{\chi_k^0}^2})
\label{new33}
\eeqn
Using the definition of their Eq.(17) there is a disagreement 
with the sign of the second part of their Eq.(16). 
Also, their $N_{4a}^*N_{a3}$ should read $N_{a4}^*N_{a3}^*$. 
The \mic group only computes $\epsilon_t^{\prime(1)}(s)$ and 
$\epsilon_t^{\prime(2)}(s)$ in our notation.  
Our approximation of the sum of these quantities gives
\beqn
\epsilon_t^{\prime(1)}(s) +\epsilon_t^{\prime(2)}(s)
=\sum_{i=1}^2\sum_{j=1}^2 \frac{2\alpha_s}{3\pi} e^{-i\xi_3}
(\mu^* +\frac{A_t}{\tan\beta}) |D_{s1i}|^2 |D_{t2j}|^2
\frac{1}{|m_{\tilde g}|} H(\frac{m_{\tilde s_i}^2}{|m_{\tilde g}|^2},
\frac{m_{\tilde t_j}^2}{|m_{\tilde g}|^2})
\label{32}
\eeqn
The analysis of the \mic group does not include CP phases. 
Setting the phases to zero; $\mu^*=\mu$, and $\xi_3=0$ etc. 
in Eq.(\ref{32}) and taking into account that they use the opposite 
sign convention for $\epsilon_t'(s)$, we find that our result is 
in complete agreement with Eq.(B.68) in Ref.\cite{Belanger:2004yn}.

\section{Numerical analysis and size estimates}
We now present a numerical analysis of our analytic 
results and also give a comparison with 
the previous works. In the following we 
compare three different methods for the calculation of the 
branching ratio of $b \rightarrow s \gamma$, by using different 
computations of the $\epsilon$'s; 
\begin{enumerate}
\item $F$: This is the full calculation of this work.
\item $S_1$: Here the $\epsilon$'s are calculated using 
the simplified formulae found in the \mic manual. 
Thus, we use the simplified formulae derived in 
Refs.\cite{Degrassi:2000qf,Pierce:1996zz,Carena:1999py} 
appropriately extended to the complex case, as derived in Sec.4. 
Moreover, we correct the neutralino mixings terms entering in 
$\epsilon_b^{\prime (3)}(t)$ and $\epsilon_b^{\prime (4)}(t)$ 
as stated in Sec.4. 
\item $S_2$: Here the $\epsilon$'s are calculated using 
the simplified formulae of Ref.\cite{Demir:2001yz}.
But with the corrections stated in this paper. 
\end{enumerate}
 In the previous section we exhibited the equivalence of our analysis and  
that of \cite{BCRS} for the  case with no CP phases, and showed that for  the 
case with CP phases our analysis is  more complete.
Thus we use our analysis in the numerical computation since it is valid
with and without CP phases.
In the numerical analysis we take the SUSY scale to be the 
average of the stop-masses.
We calculate the difference in percent to the full 
$b \rightarrow s \gamma$ calculation via the relation
\begin{equation}
  \frac{{\rm BR}(b \rightarrow s \gamma)_F-{\rm BR}(b \rightarrow s \gamma)_S}
{{\rm BR}(b \rightarrow s \gamma)_F}
\end{equation}
where $S=S_1,S_2$. 
 In our numerical analysis we investigate several different supersymmetry  
breaking scenarios. These are
\begin{enumerate}
\item 
 mSUGRA with real soft terms and complex SUGRA with universal 
value for the absolute soft terms.
\item
MSSM with real and complex soft breaking sector.
\end{enumerate}
In the analysis we scan over
the parameter-space in order to find the qualitatively difference 
among the schemes above  and search for the parameter space  where the 
   allowed points satisfy the experimentally measured rate 
for $b \rightarrow s \gamma$ within 2 $\sigma$, thus requiring
\begin{equation}
   2.3 \times 10^{-4} < {\rm BR}(b \rightarrow s \gamma)_F 
   < 4.7 \times 10^{-4}
\label{bounds}
\end{equation}
In addition we check that all bounds on sparticle 
masses are satisfied, where we use the bounds given in Ref.\cite{pdg}.
Furthermore, we require the Higgs mass to be heavier than 110 GeV in the 
real case, as the theoretical error in the calculation of its mass 
is of order a few GeV. In the complex case we choose the lower 
bound 100 GeV for the lightest Higgs mass, as in this case there 
is a possibility for such a low mass being consistent with the 
LEP data~\cite{carena:2000ks}. This choice has little influence on 
our results. 
 For the computation of the Higgs mass we use 
{\tt CPsuperH}~\cite{Lee:2003nt}. 

Clearly, some of the contributions in the $\epsilon$'s are 
numerically insignificant. We find that in $\epsilon_{bb}$ 
the contributions $\epsilon_{bb}^{(5)}$ and $\epsilon_{bb}^{(6)}$ 
are small. Also the contributions  $\epsilon^{\prime(3),(4)}_t(s)$ can be 
safely neglected, as the terms that would have been dominating 
are suppressed by the strange-quark Yukawa coupling. 
However, the contribution $\epsilon^{\prime(5),(6)}_t(s)$, which has not been
 included in  $S_1$ and $S_2$ calculations of the rate for 
$b \rightarrow s \gamma$ gives sizeable contribution, capable of 
changing the rate by a few percent. 
 The CKM elements $V_{qq'}$ that enter the analysis above are radiatively
corrected and are calculated following the work of \cite{babunew}. Numerically
the radiative corrections are found to be small in the part
of parameter space investigated but the corrections could be significant
in other parts of the parameter space.

Before proceeding further we exhibit the dependence of the $\epsilon$'s
and the $b\rightarrow s\gamma$ branching ratio on phases. This is done
in Fig. (5) which shows sharp dependence of these quantities on the phases.
Specifically the analysis ob $b\rightarrow s\gamma$ in Fig. (5) shows that the
effect of phases can move the branching ratio for a given point in
the parameter space from the experemintally forbidden area into
the allowed corridor of values in Eq.(\ref{bounds}).

In the following we discuss the different scenarios in detail.

\subsection{mSUGRA and complex SUGRA with universalities}
In this section we carry out an analysis in the 
framework of extended mSUGRA model 
whose soft breaking sector is described by the parameters
$m_0, A_0 =|A_0|e^{i\alpha}, \tan\beta, \mu_0=|\mu_0|e^{i\theta_{\mu}}, 
\tilde m_i=|m_{\frac{1}{2}}|e^{i\xi_i}$ (i=1,2,3) 
where $m_0$ is the universal scalar mass, $m_{\frac{1}{2}}$ 
is the universal gaugino mass, $A_0$ is the universal trilinear coupling,
and $\mu_0$ is the  Higgs mixing parameter, while 
$\alpha, \theta_{\mu}, \xi_i$ are the phases, all taken at the 
GUT scale. 

 In the real mSUGRA case the scan is done by randomly selecting 
points within the following 
parameter-space; $m_0 \in [200,1000]$ GeV, $m_{1/2} \in [200,1000]$ GeV, 
$A_0 \in [-3 m_0,3 m_0]$, $\tan\beta \in [5,55]$ and both signs 
of the $\mu$ parameter. For the complex case we also vary the 
five phases $\theta_\mu, \alpha_A, \xi_1,\xi_2, \xi_3$ within 
the range zero to $\pi$. 

Our results are shown in Fig.\ref{bsgsugra}. 
We find a significant correlation of the increase of the 
differences with $\tan\beta$. 
This is very natural as the physical important parameter is 
$\epsilon \tan\beta$ compared to one. 
Thus, in order for the $\epsilon$ parameters to have a 
substantial influence $\tan\beta$ must be large.
We see that in both cases the \mic approximation 
is better (once we include the appropriate phases on their
expressions) for the real case the differences remain below 
2\% using method $S_1$ and for method $S_2$ it is 
less than about 4\%. 
In the complex case while the $S_1$ approximation remain below 
4\%, the differences in the $S_2$ approximation can reach 8\%. 
This can be attributed to the fact that $S_2$ does not include 
the electroweak contributions to $\epsilon_{bb}$ and 
$\epsilon_b^\prime(t)$. 
These contributions can induce a relatively large error at 
small values of $m_{1/2}$ (indeed all the points with $S_2$ 
$\sim$ 8\% correspond to $m_0<400$ GeV and $m_{1/2}<250$ GeV). 
These results can be applied also to the supersymmetric 
corrections to the b-quark mass, $\delta m_b$, which is 
given by $\epsilon_{bb}\tan\beta$.  
We find that in the mSUGRA and complex SUGRA cases 
the simplification of $S_2$ provides an accuracy of about 40\% and 
the simplification of $S_1$ an accuracy of about 5\%.

We would like to stress the importance of using the correct 
signs and complex-conjugates. In Fig.\ref{bsgsugrawrong} we 
compare again the methods $S_1$ and $S_2$ against our 
full calculations, but this time we use the original 
formulae, as presented
in Ref.\cite{Demir:2001yz} and Ref.\cite{Belanger:2004yn}
\footnote{We note that the calculation in 
\mic, is indeed performed correctly, despite of the 
error in the analytical formula. This is due to the fact that 
they use the a real $N$ and thus allow for negative 
mass-eigenvalues, which is numerically a valid procedure in case of 
real parameters.}.
The difference is seen to increase substantial being 
as much as 15\%. This can be understood since often there 
are cancellations among the epsilons contribution to the 
Wilson Coefficients. For instance the 
SUSY corrections to the W-contribution scale with 
$\epsilon_{bb}^*-\epsilon_{b}^{\prime}(t)$ and this factor 
is much smaller than the $\epsilon$'s themselves due to  
cancellation. Thus, having a wrong sign on one of the terms  
can cause a large effect on the SUSY correction.

To compare the accuracy of the various methods of evaluation of the 
$\epsilon$'s we focus on the point:
\beq
m_0=350 ,\; \; m_{1/2}=200 ,\; \; \; A_{0}=700,\; \; \;
\xi_1=0,\; \; \; \xi_2=0.75,\; \; \;  
\xi_3=0.5,\; \; \; \theta_\mu=0.6 \; \; \; \alpha=\pi \;,
\label{point0}
\eeq
where all the phases are given in radians and masses in GeV. 
Our results are shown in Fig.~\ref{pointsugra}. 
For method $S_1$ the differences can be attributed to the 
simplification of the calculations of some terms 
and to neglecting the terms $\epsilon_t^{\prime (5)}(s)$ 
and $\epsilon_t^{\prime (6)}(s)$. 
We note that the inclusion of the electroweak contributions 
in $\epsilon_{bb}$ and $\epsilon_{b}^\prime(t)$,  
is an improvement as compared to the simplified $S_2$ method.

Overall the corrections given in this paper to the 
BR($\bsg$) are relatively small in the SUGRA scenario. 
Clearly, the SUGRA scenario constrains the MSSM mass 
spectra to have certain hierarchies. Moreover, due to the 
RGE evolution in SUGRA one normally finds the low-energy 
trilinear top term to be $A_t \sim - M_3$, unless the 
GUT scale $A_0$ is very large.
As we now discuss, these constraints in the SUGRA 
scenario give rise to various cancellations. 
The phase of the LO chargino contribution (see Eq.(\ref{chargino})), 
assuming the hierarchy imposed in SUGRA scenarios,  is given by 
Arg(${\mu A_t}$) and the phase of the NLO chargino contribution 
is Arg(${-\mu M_3}$). The LO Higgsino contribution as well as 
W contribution are always positive. Thus, for the Higgsino and 
the chargino contribution to cancel against each other one needs 
Arg(${\mu A_t}) \sim \pi$. If such a cancellation occurs  
the SUSY corrections are allowed to be large. 
As noted in Sec.3 all SUSY corrections scales with 
$1/(1+\epsilon_{bb}^*\tan\beta)$ and thus these corrections 
will be large if $\epsilon_{bb}^*\tan\beta$ is close to minus one. 
The leading SUSY QCD contribution to $\epsilon_{bb}$ has a 
phase of Arg($\mu M_3$). And this term is positive in the case of 
a negative chargino contribution. 
Thus, in mSUGRA one can never have a cancellation 
between the Chargino and the Higgsino contributions 
and at the same time have a negative value of $\epsilon_{bb}$. 
This is the main reason that the differences in the SUGRA case 
are rather small. 
Another reason is that in general in the SUGRA scenario 
the different contributions to the $\epsilon$'s cancels 
against each other. 
Thus, for instance for the $\epsilon_{bb}$ correction, the 
leading SUSY QCD correction has the opposite sign as compared 
to the Yukawa and the electroweak contributions. Again this 
cancellation arises due to the relation between the trilinear 
top term and the gluino mass. 
Thus, in the SUGRA scenario the $\epsilon$'s are numerically 
smaller than in the general MSSM scenario.

\subsection{MSSM with real and complex soft breaking sector}

In the MSSM case the scan is done by randomly choosing the 
soft-masses in the range 200 to 1000 GeV and the trilinear terms 
between $-3000,3000$. We also run with plus and minus sign on the 
$\mu$-term, the trilinear terms and the gaugino masses. 
In the complex case we take the phases between 0 to $\pi$. 
We take the first  and the second generation squarks and sleptons 
to be degenerate and we parameterize their masses by $m_{sl}$. 
In the CP-violating case we use fairly heavy first and 
second generation, $m_{sl}=2000$ GeV, in order not to generate 
large EDM's. The first and second generations do not influence our 
calculation a lot, but they do enter in the evaluation of the 
chargino contribution to the Wilson coefficients. 
They also enter in the evaluation of 
$\epsilon^{\prime}_t(s)$, but only in the terms  
$\epsilon^{\prime (3)}_t(s)$ and $\epsilon^{\prime(4)}_t(s)$, 
that can be safely neglected as they are numerically very small.

In the MSSM case Fig.\ref{bsgmssm} shows that the difference 
using method $S_1$ can be as large as 60\% and for method $S_2$ 
we find roughly the same upper bound but the average difference 
is larger. 
Although on average the simplified 
formulae do a good job, there are cases with large errors
in the rate for $b \rightarrow s \gamma$.  
The large difference occurs  in the MSSM case, 
particularly when there is a large sbottom mixing 
and/or large stop mixing. 
It is not difficult to realize this by looking at the formulae
for the $\epsilon$'s. 
Looking at the simplified formula for $\epsilon_{bb}^{(1)}$, 
one notice that these neglect the sbottom mixing, as 
compared to the full formulae.  
However, in the limit that the sbottom masses are equal the 
dominant term in the full formulae is invariant under sbottom 
mixing. Thus, in order for this correction to be important 
one needs a large sbottom mixing and a large sbottom mass difference, 
which occurs 'rarely'. 
As an example, in the formula for $\epsilon^{\prime (1)}_b(t)$, 
the first term (neglected in previous works) is 
\begin{equation}
-\sum_{i=1}^2\sum_{j=1}^2 \frac{2\alpha_s}{3\pi} e^{i\xi_3}
\frac{m_t}{m_b} \cot\beta  A_{t} D_{b1j} D^{*}_{t2i}
D^{*}_{b2j}D_{t1i}
\frac{1}{|m_{\tilde g}|} H(\frac{m_{\tilde t_i}^2}{|m_{\tilde g}|^2}, 
\frac{m_{\tilde b_j}^2}{|m_{\tilde g}|^2})
\end{equation}
The factor $m_t/m_b$ easily overcomes the suppression by $\cot\beta$. 
However, this term is zero in the limit of no sbottom or stop mixing.
But, when the sbottom and stop mixing is large it 
can contribute significantly. Even the term with 
$m_t^2/m_b \cot\beta D_{b1j}D_{t1i}^*$ can give non-negligible 
contribution. For point (i) defined in Table.\ref{points} we 
have the particular 
situation of both large stop as well as large sbottom mixing. 
We show the values of the $\epsilon$'s and the rate of 
$b \rightarrow s \gamma$ in Fig.(\ref{point1plot}).
It is seen that in particular the value of $\epsilon_b^{\prime}(t)$
is deviating from the value of the simplified formulae. 
Notice that this point with $\mu$ negative is excluded with the 
simplified formulae, but allowed with the full calculation.  
The opposite might also occur as shown in Fig.(\ref{point2plot}).
 
The phase of the chargino contribution in the MSSM case depends 
on the mass hierarchies of the sparticles, and is thus less 
restricted than in the SUGRA scenario. Also, in the MSSM 
we are no longer confined to have $\alpha \sim \pi-\xi_3$. 
Therefore, it is possible to have points where the chargino and 
the 
Higgsino contribution have opposite signs and at the same 
time have $\epsilon_{bb}^*\tan\beta$ negative. 
Indeed all points with a difference of more than 
20\%, assuming real parameters, have a value of 
$\epsilon_{bb}\tan\beta$ less than minus one half.
For the point(i) plotted in Fig.\ref{point1plot}, the Chargino and 
Higgsino contribution are even larger than the SM contribution 
but as they cancels against each other, one finds a BR($\bsg$) in 
agreement with experiment. 

\begin{table}[h]
\begin{center}
\begin{tabular}{|r|r|r|r|r|r|r|r|r|r|r|r|r|r|}
\hline   
Point & $M_H^{+}$ & $\mu$ & $M_1$ & $M_2$ & $M_3$ &
$M_{\tilde{Q}}$ & $M_{\tilde{U}}$ & $M_{\tilde{D}}$ & $M_{\tilde{L}}$ &
$M_{\tilde{E}}$ & $A_t$ & $A_b$ & $A_\tau$  \\ \hline       
(i)  & 450  & -950  & 200  & -300 & 400  & 950  &
900  & 700  &  300  & 300  & 2500 & 1000 & 0 \\ \hline
(ii)  & 300  & 700  & 200  &  300 & -700  & 550  &
600  & 500  &  300  & 300  & -1000 &  0 & 0 \\ \hline
\end{tabular}
\end{center}
\caption{Values of the parameters for point (i) and point (ii) in GeV. 
The value of $\tan\beta$ is not fixed and in both cases we have 
used $m_{sl}=500$ GeV.}
\label{points}
\end{table}

\section{Conclusion}
\begin{figure}[t]
\hspace*{-0.8in}
\centering
\includegraphics[width=20cm,height=20cm]{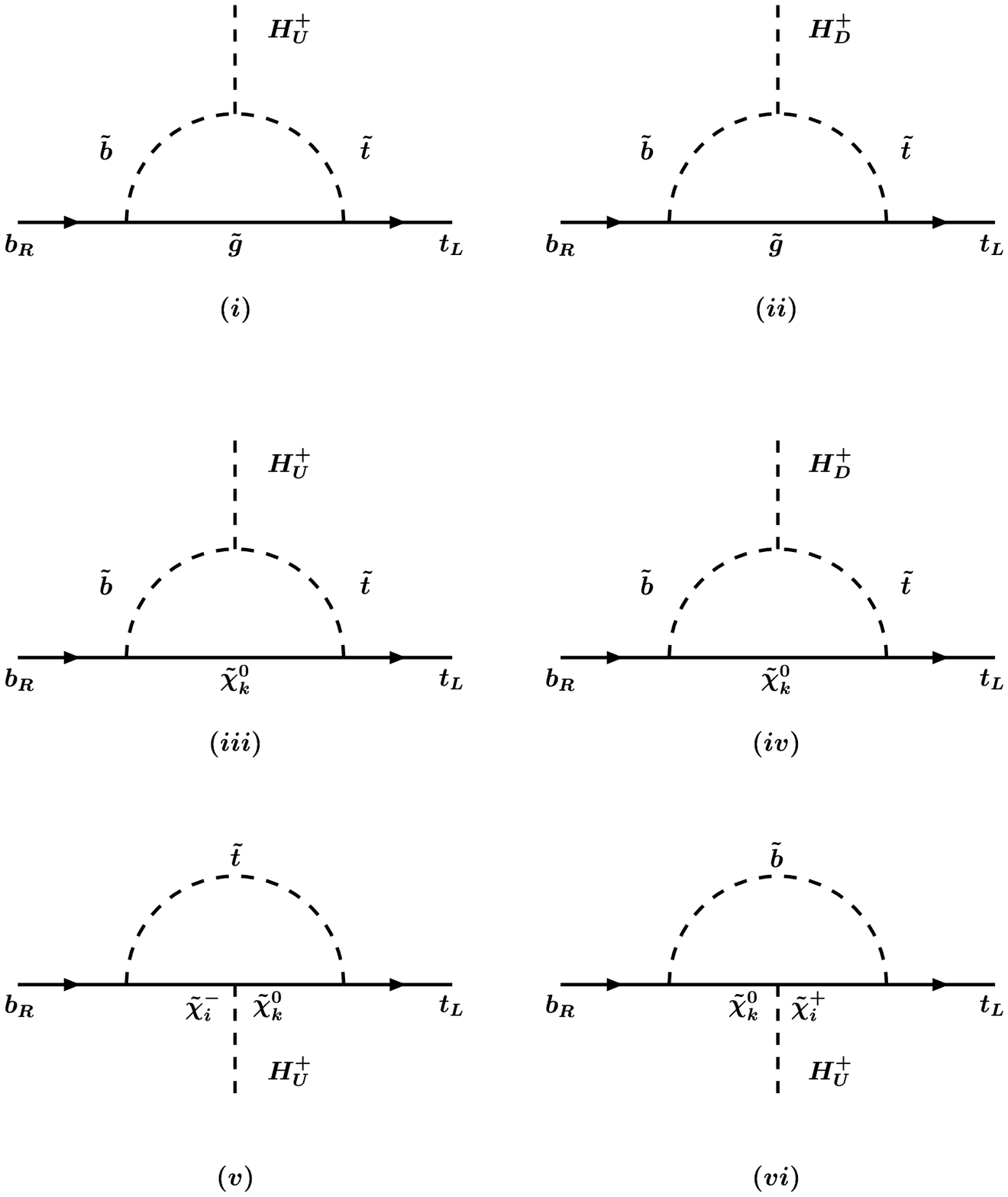}
\caption{ Set of diagrams contributing to $\epsilon_b'(t)$  }
\label{epsfig1}
\end{figure}

\begin{figure}[t]
\hspace*{-0.8in}
\centering
\includegraphics[width=20cm,height=20cm]{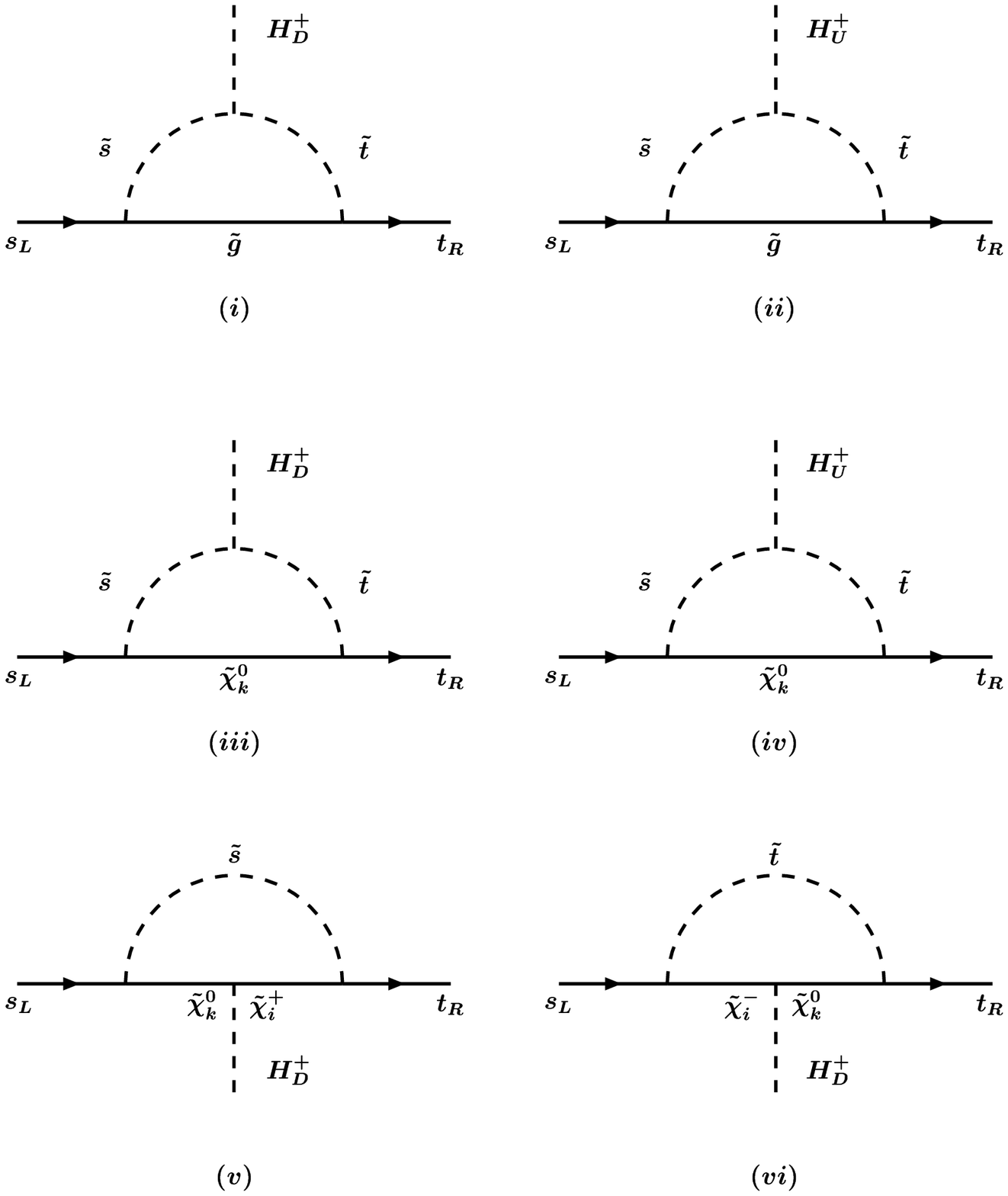}
\caption{ Set of diagrams contributing to $\epsilon_t'(s)$  }
\label{epsfig2}
\end{figure}

\begin{figure}[t]
\hspace*{-0.8in}
\centering
\includegraphics[width=20cm,height=20cm]{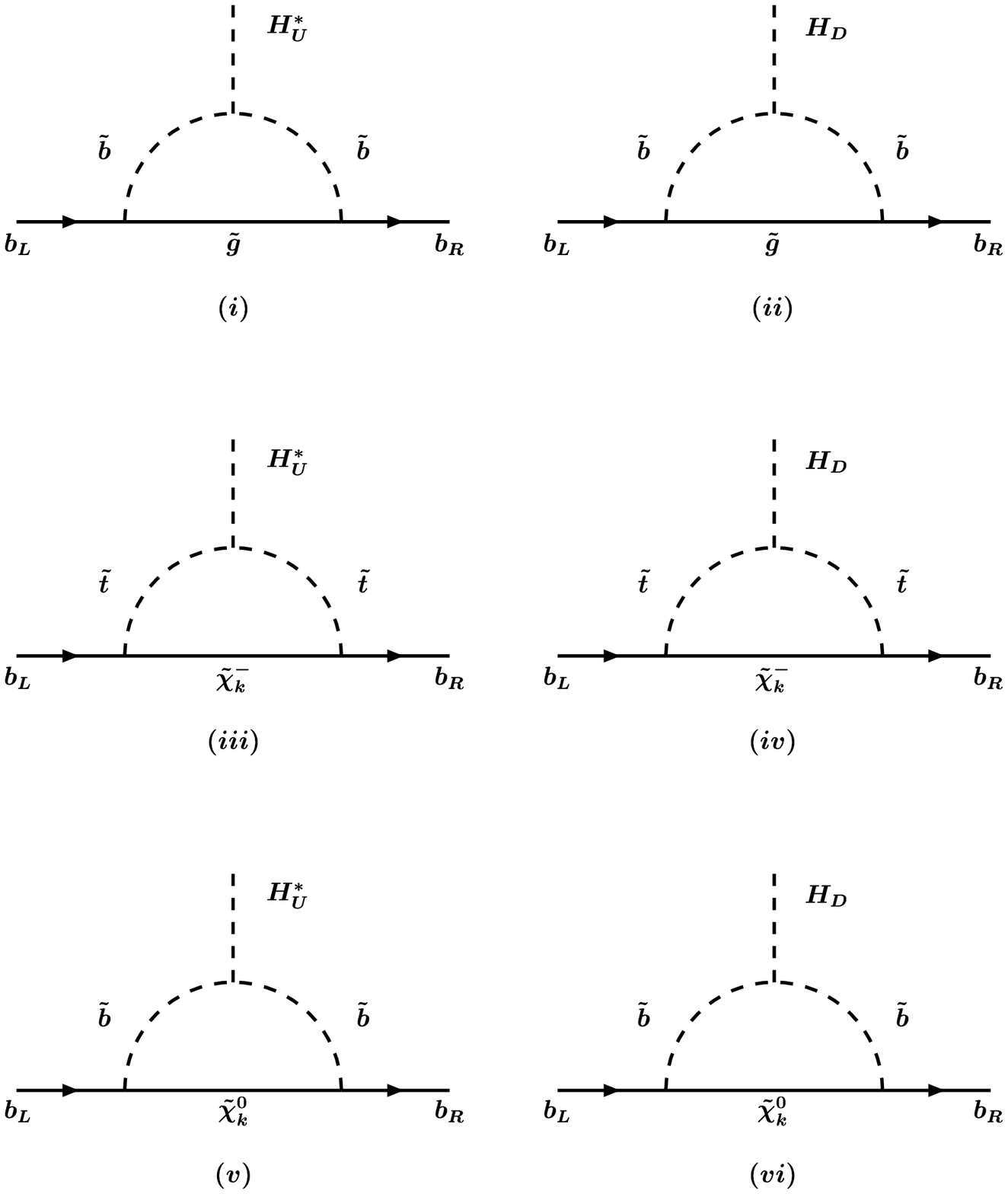}
\caption{ Set of diagrams contributing to $\epsilon_{bb}$}
\label{epsfig3}
\end{figure}

\begin{figure}[t]
\hspace*{-0.8in}
\centering
\includegraphics[width=20cm,height=20cm]{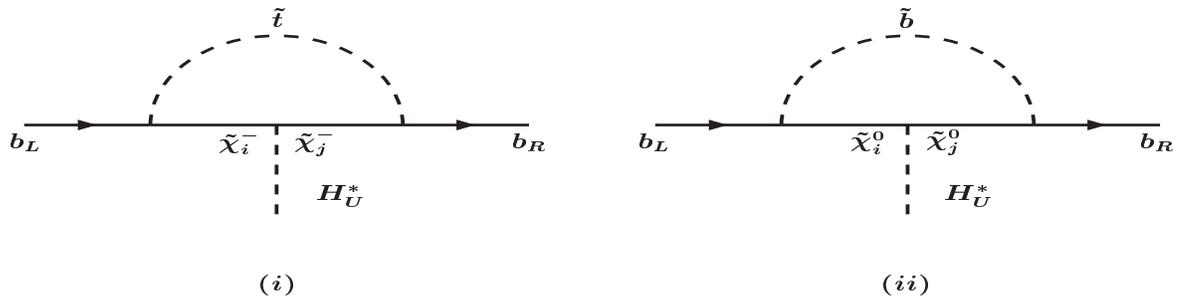}
\caption{ Additional diagrams contributing to $\epsilon_{bb}$}
\label{epsfig4}
\end{figure}

In this paper we have given a more complete analysis of the 
next-to-leading-order contributions which are enhanced by 
$\tan\beta$ factors. Such corrections affect the Wilson coefficients
$C_7$ and $C_8$ arising from the $W$, Higgs $H^{\pm}$, and chargino
$\chi^{\pm}$ exchange contributions.  There are twenty supersymmetric 
one-loop diagrams that contribute to these corrections. Some of these
loops have been computed in previous works. In this paper we 
have given an analytic analysis of the full set of these 
corrections which involves computations of the  
six diagrams of Fig.\ref{epsfig1}, Fig.\ref{epsfig2} and 
Fig.\ref{epsfig3} each and the two diagrams of Fig.\ref{epsfig4}. 
The analysis presented here also includes the full CP phase
dependence allowed within the general soft breaking in MSSM. 
The new analytic results of this paper are contained in 
Appendices B,C and D. In Sec.4 we gave a comparison of the 
current work with previous analyses.
In Sec.5 a numerical analysis of the corrections was given and 
the effect of corrections found to be significant 
specifically when there are large sbottom and stop mixings in the 
general MSSM case. The vertex corrections 
derived in this paper are relevant for a variety of 
phenomena where sparticles enter in the loops or are 
directly produced in the laboratory, such 
as Higgs decay widths and lifetimes and for the 
supersymmetric corrections to the b-quark mass. 
 Since the analysis presented here takes full account of the 
effect of CP phases on $b\rightarrow s+\gamma$, 
it should serve as an important tool for testing supersymmetric models.

\noindent
{\large\bf Acknowledgments}\\ 
MEG acknowledges support from the 'Consejer\'{\i}a de Educaci\'on de 
la Junta de Andaluc\'{\i}a', the Spanish DGICYT under contract 
BFM2003-01266 and European Network for Theoretical Astroparticle 
Physics (ENTApP), member of ILIAS, EC contract number
RII-CT-2004-506222. 
TI acknowledges the hospitality extended to him by the American 
University of Beirut.
The research of TI and PN was supported in part by NSF grant PHY-0546568. 
SS is supported by Fundac\~{a}o de Amparo \`{a} Pesquisa do
Estado de S\~{a}o Paulo (FAPESP).

\noindent
{\bf Appendix A}\\
Here we list the numerical values of the coefficients 
$h_i, \bar h_i, a_i, b_i$ that appear in Eq.(8)\cite{Chetyrkin:1996vx}. 
\beqn
h_i &=& ( \frac{626126}{272277}, -\frac{56281}{51730}, 
-\frac{3}{7}, -\frac{1}{14}, -0.6494, 
-0.0380, -0.0186, -0.0057)\nonumber\\
a_i &=& (\frac{14}{23}, \frac{16}{23},\frac{6}{23}, 
-\frac{12}{23},  0.04086, -0.4230, -0.8994, 0.1456) \nonumber\\
\bar h_i &=& (-0.9135, 0.0873, -0.0571, 0.0209) \nonumber\\
b_i &=& (0.4086, -0.4230, -0.8994, 0.1456)
\eeqn 

\noindent
{\bf Appendix B - Analysis of $\epsilon_b'(t)$}\\

The analysis of $\epsilon_b'(t)$ as well as of $\epsilon_t'(s)$ 
and of $\epsilon_{bb}$ depends on the soft breaking parameters. 
We shall carry out the analysis in the framework of MSSM
which has a pair of Higgs doublets with Higgs mixing parameter 
$\mu$ which is in general complex, assuming a general set of 
soft breaking parameters. Specifically we will assume 
for the $\epsilon$ analysis a general set of squark masses, 
and of trilinear couplings $A_q$ which we assume in general to be complex. 
Similarly, we assume the gaugino masses $\tilde m_i$ (i=1,2,3) to be complex.
Thus our analytic analysis will not be tied to any specific model of 
soft breaking.  There are six different loop diagrams that contribute to
$\epsilon_b'(t)$ so that 
\beqn
\epsilon_b'(t) = \sum_{i=1}^6 \epsilon_b^{\prime(i)}(t)  
\eeqn
We exhibit now each of the above contributions.\\ 
From Fig.\ref{epsfig1}(i) we find 
\beqn
\epsilon_b^{\prime(1)}(t)&=& 
-\sum_{i=1}^2\sum_{j=1}^2 \frac{2\alpha_s}{3\pi} e^{i\xi_3}D_{b2j}^*D_{t1i} 
[\frac{m_t}{m_b} \cot\beta A_t D_{b1j}D_{t2i}^*+ \mu D_{b2j}D_{t1i}^*
+ m_t \cot\beta D_{b2j}D_{t2i}^*  \nonumber\\
&& +\frac{m_t^2}{m_b} \cot\beta D_{b1j}D_{t1i}^* -\frac{m_W^2}{m_b} 
\sin\beta \cos\beta D_{b1j}D_{t1i}^* ]
\frac{1}{|m_{\tilde g}|} H(\frac{m_{\tilde t_i}^2}{|m_{\tilde g}|^2}, 
\frac{m_{\tilde b_j}^2}{|m_{\tilde g}|^2})
\eeqn
where $D_q$ is the matrix that diagonalizes the squark mass-squared 
matrix, i.e., 
\beqn
D_q^{\dagger} M_{\tilde q}^2D_q ={\rm diag} 
(M_{\tilde q_1}^2, M_{\tilde q_2}^2)
\eeqn
\noindent
From Fig.\ref{epsfig1}(ii) we find 
\beqn
\epsilon_b^{\prime(2)}(t) &=& 
\frac{1}{\tan\beta} \sum_{i=1}^2\sum_{j=1}^2 \frac{2\alpha_s}{3\pi} e^{i\xi_3}
D_{b2j}^*D_{t1i} 
[  A_b^* D_{b2j}D_{t1i}^*  +  \frac{m_t}{m_b} \mu^* \cot\beta D_{b1j}D_{t2i}^*
 +m_t  D_{b2j}D_{t2i}^* \nonumber \\
&& + m_b  D_{b1j}D_{t1i}^*
-\frac{m_W^2}{m_b} \cos^2\beta D_{b1j}D_{t1i}^* ]
\frac{1}{|m_{\tilde g}|} H(\frac{m_{\tilde t_i}^2}{|m_{\tilde g}|^2}, 
\frac{m_{\tilde b_j}^2}{|m_{\tilde g}|^2}) 
\eeqn
This diagram is not enhanced by $\tan\beta$. 
From Fig.\ref{epsfig1}(iii) we find 
\beqn
\epsilon_b^{\prime(3)}(t)&=&2 \sum_{k=1}^4\sum_{i=1}^2\sum_{j=1}^2 
[\frac{m_t}{m_b} \cot\beta A_t D_{b1j}D_{t2i}^*+ \mu D_{b2j}D_{t1i}^* \nonumber\\
&& +m_t \cot\beta D_{b2j}D_{t2i}^*
+\frac{m_t^2}{m_b} \cot\beta D_{b1j}D_{t1i}^* 
-\frac{m_W^2}{m_b} \sin\beta \cos\beta D_{b1j}D_{t1i}^* ]\nonumber\\
&& (\alpha_{bk}^* D_{b1j}^* - \gamma_{bk}^*D_{b2j}^*)
(\beta_{tk}D_{t1i}+\alpha_{tk}^* D_{t2i})
\frac{1}{16\pi^2} 
\frac{1}{m_{\chi_k^0}} H(\frac{m_{\tilde t_i}^2}{m_{ \chi_k^0}^2}, 
\frac{m_{\tilde b_j}^2}{m_{\chi_k^0}^2})
\eeqn
In the above  $\alpha$, $\beta$,  and $\gamma$ for the b and t quarks are defined 
so that
\beqn\label{alphabk}
\alpha_{b(t)k} &=& \frac{g
  m_{b(t)}X_{3(4)k}}{2m_W\cos\beta(\sin\beta)} \nonumber \\ 
\beta_{b(t)k} &=& eQ_{b(t)}X_{1k}^{'*} +\frac{g}{\cos\theta_W} X_{2k}^{'*}
(T_{3b(t)}-Q_{b(t)}\sin^2\theta_W)\nonumber\\ 
\gamma_{b(t)k} &=& eQ_{b(t)} X_{1k}'-\frac{gQ_{b(t)}\sin^2\theta_W}{\cos\theta_W}
X_{2k}'
\eeqn
where $Q_{b(t)}= -\frac{1}{3}(\frac{2}{3})$ 
and $T_{3b(t)}=-\frac{1}{2}(\frac{1}{2})$ and where 
\begin{equation}
X'_{1k} = X_{1k}\cos\theta_W +X_{2k}\sin\theta_W \;\;,\;\;
X'_{2k} = -X_{1k}\sin\theta_W +X_{2k}\cos\theta_W
\end{equation}
and $X$ diagonalizes the neutralino mass matrix. 
\beqn
~X^T M_{\chi^0}X= {\rm diag} (m_{\chi_1^0}, m_{\chi_2^0},m_{\chi_3^0},m_{\chi_4^0}) 
\eeqn

\noindent
From Fig.\ref{epsfig1}(iv),  which is non-$\tan\beta$ enhanced, we find 
\beqn
\epsilon_b^{\prime(4)}(t) &=& -\frac{2}{\tan\beta} \sum_{k=1}^4
 \sum_{i=1}^2\sum_{j=1}^2 
 [  A_b^* D_{b2j}D_{t1i}^*  +  \frac{m_t}{m_b} \mu^* \cot\beta
 D_{b1j}D_{t2i}^*  \nonumber\\
&& + m_t  D_{b2j}D_{t2i}^* + m_b  D_{b1j}D_{t1i}^*
-\frac{m_W^2}{m_b} \cos^2\beta D_{b1j}D_{t1i}^* ] \nonumber \\
&& (\alpha_{bk}^*D_{b1j}^*-\gamma_{bk}^*D_{b2j}^*)
(\beta_{tk}D_{t1i}+\alpha_{tk}^*D_{t2i}) 
\frac{1}{16\pi^2} 
\frac{1}{m_{\chi_k^0}} H(\frac{m_{\tilde t_i}^2}{m_{\chi_k^0}^2}, 
\frac{m_{\tilde b_j}^2}{m_{\chi_k^0}^2})
\eeqn
\noindent
From Fig.\ref{epsfig1}(v) we find 
\beqn
\label{epsb5}
\epsilon_b^{\prime(5)}(t) &=& \sum_{k=1}^4\sum_{i=1}^2\sum_{j=1}^2 
\sqrt 2 g \frac{m_W}{m_b} \cos\beta [K_b U_{i2} D_{t1j}^*] 
  ~ (\sqrt 2 X_{4k}^* V_{i1} +X_{2k}^* V_{i2} +\tan\theta_W X_{1k}^* V_{i2}) 
\nonumber\\
&& \frac{1}{16\pi^2} 
\frac{m_{\chi_i^-}m_{\chi_k^0}}{ m_{\tilde t_j}^2}
(\beta_{tk}D_{t1j}+\alpha_{tk}^*D_{t2j}) 
 H(\frac{m_{ \chi_i^-}^2}{m_{\tilde t_j}^2} , 
\frac{m_{\chi_k^0}^2}{m_{\tilde t_j}^2} )          
\eeqn
In the above $U$ and $V$ are the matrices that diagonalize the chargino mass 
matrix 
\beqn
U^*M_{\chi^+}V^{-1} ={\rm diag} ( m_{\chi_1^+}, m_{\chi_2^+})
\eeqn
and $K_b$ is given by 
\beqn
K_b=\frac{m_b}{\sqrt 2 m_W \cos\beta}
\eeqn
Finally, from Fig.\ref{epsfig1}(vi) we get 
\beqn
\label{epsb6}
\epsilon_b^{\prime(6)}(t) &=& -\sum_{k=1}^4\sum_{i=1}^2\sum_{j=1}^2 
\sqrt 2 g \frac{m_W}{m_b} \cos\beta (U_{i1}D_{b1j}-K_b U_{i2}D_{b2j}) 
\nonumber\\
&& (\sqrt 2 X_{4k}^* V_{i1} +X_{2k}^* V_{i2} +\tan\theta_W X_{1k}^* V_{i2}) 
\nonumber\\
&& \frac{1}{16\pi^2} 
\frac{m_{\chi_i^-}m_{\chi_k^0}}{ m_{\tilde b_j^2}}
(\alpha_{bk}^*  D_{b1j}^*-\gamma_{bk}^*D_{b2j}^*) 
 H(\frac{m_{ \chi_i^-}^2}{m_{\tilde b_j}^2}, 
\frac{m_{\chi_k^0}^2}{m_{\tilde b_j}^2} )          
\eeqn
%
\noindent
{\bf Appendix C - Analysis of $\epsilon_t'(s)$}\\

Next we look at the $\epsilon_t'(s)$ analysis. Here we have 
\beqn
\epsilon_t'(s) = \sum_{i=1}^6 \epsilon_t^{\prime(i)}(s)  
\eeqn
The individual contributions $\epsilon_t^{\prime(i)}(s)$ are 
exhibited below.

\noindent
From Fig.\ref{epsfig2}(i) we find 
\beqn
\epsilon_t^{\prime(1)}(s) &=&
\sum_{i=1}^2\sum_{j=1}^2 \frac{2\alpha_s}{3\pi} e^{-i\xi_3}D_{s1i}^*D_{t2j} 
[\frac{m_s}{m_t} \tan\beta A_s^* D_{s2i}D_{t1j}^*+ \mu^*
D_{s1i}D_{t2j}^* + m_s \tan\beta D_{s2i}D_{t2j}^* \nonumber\\
&& +\frac{m_s^2}{m_t} \tan\beta D_{s1i}D_{t1j}^* 
-\frac{m_W^2}{m_t} \sin\beta \cos\beta D_{s1i}D_{t1j}^* ]
\frac{1}{|m_{\tilde g}|} H(\frac{m_{\tilde s_i}^2}{|m_{\tilde g}|^2}, 
\frac{m_{\tilde t_j}^2}{|m_{\tilde g}|^2})
\eeqn

\noindent
From Fig.\ref{epsfig2}(ii), which is non-$\tan\beta$ enhanced, we find 
\beqn
\epsilon_t^{\prime(2)}(s)=
\frac{1}{\tan\beta} \sum_{i=1}^2\sum_{j=1}^2 \frac{2\alpha_s}{3\pi} e^{-i\xi_3}
D_{s1i}^*D_{t2j} 
[  A_t D_{s1i}D_{t2j}^*  +  \frac{m_s}{m_t} \mu \tan\beta D_{s2i}D_{t1j}^*
\nonumber\\
+m_s  D_{s2i}D_{t2j}^* 
+ m_t  D_{s1i}D_{t1j}^*
-\frac{m_W^2}{m_t} \sin^2\beta D_{s1i}D_{t1j}^* ]
\frac{1}{|m_{\tilde g}|} H(\frac{m_{\tilde s_i}^2}{|m_{\tilde g}|^2}, 
\frac{m_{\tilde t_j}^2}{|m_{\tilde g}|^2})
\eeqn

\noindent
From Fig.\ref{epsfig2}(iii) we find 
\beqn
\epsilon_t^{\prime(3)}(s) &=& -2 \sum_{k=1}^4\sum_{i=1}^2\sum_{j=1}^2 
[\frac{m_s}{m_t} \tan\beta A_s^* D_{s2i}D_{t1j}^*+ \mu^*
D_{s1i}D_{t2j}^* + m_s \tan\beta D_{s2i}D_{t2j}^* \nonumber\\
&& +\frac{m_s^2}{m_t} \tan\beta D_{s1i}D_{t1j}^* 
-\frac{m_W^2}{m_t} \sin\beta \cos\beta D_{s1i}D_{t1j}^* ] \nonumber\\
&& (\beta_{sk}^* D_{s1i}^*+\alpha_{sk} D_{s2i}^*)
(\alpha_{tk} D_{t1j} - \gamma_{tk} D_{t2j} )
\frac{1}{16\pi^2} 
\frac{1}{m_{\chi_k^0}} H(\frac{m_{\tilde s_i}^2}{m_{ \chi_k^0}^2}, 
\frac{m_{\tilde t_j}^2}{m_{\chi_k^0}^2})
\eeqn

\noindent
From Fig.\ref{epsfig2}(iv) we find  the non-$\tan\beta$ enhanced 
contribution
\beqn
\epsilon_t^{\prime(4)}(s) &=& -\frac{2}{\tan\beta} \sum_{k=1}^4
 \sum_{i=1}^2\sum_{j=1}^2 
 [  A_t D_{s1i}D_{t2j}^*  +  \frac{m_s}{m_t} \mu \tan\beta D_{s2i}D_{t1j}^* 
 \nonumber\\
&&+m_s  D_{s2i}D_{t2j}^* + m_t  D_{s1i}D_{t1j}^*
-\frac{m_W^2}{m_t} \sin^2\beta D_{s1i}D_{t1j}^* ]\nonumber\\
&& (\alpha_{tk} D_{t1j} - \gamma_{tk} D_{t2j} )
(\beta_{sk}^* D_{s1i}^*+\alpha_{sk} D_{s2i}^*)
\frac{1}{16\pi^2} 
\frac{1}{m_{\chi_k^0}} H(\frac{m_{\tilde s_i}^2}{m_{ \chi_k^0}^2}, 
\frac{m_{\tilde t_j}^2}{m_{\chi_k^0}^2})
\eeqn
\noindent
From Fig.\ref{epsfig2}(v) we find 
\beqn
\epsilon_t^{\prime(5)}(s)&=& \sum_{k=1}^4\sum_{i=1}^2\sum_{j=1}^2 
\sqrt 2 g \frac{m_W}{m_t} \sin\beta [K_t V_{i2}^* D_{s1j}]
\nonumber\\
&& (-\sqrt 2 X_{3k} U_{i1}^* +X_{2k} U_{i2}^* +\tan\theta_W X_{1k} U_{i2}^*)  
\nonumber\\
&& (\beta_{sk}^*  D_{s1j^*}+\alpha_{sk}D_{s2j}^*) 
\frac{1}{16\pi^2} 
\frac{m_{\chi_k^0}m_{\chi_i^-}}{ m_{\tilde s_j^2}}
 H(\frac{m_{ \chi_i^-}^2}{m_{\tilde s_j}^2} , 
\frac{m_{\chi_k^0}^2}{m_{\tilde s_j}^2} )          
\eeqn
where
\beqn
K_{t}=\frac{m_{t}}{\sqrt 2 m_W\sin\beta}
\eeqn
\noindent
From Fig.\ref{epsfig2}(vi) we find 
\beqn
\epsilon_t^{\prime(6)}(s) &=& - \sum_{k=1}^4\sum_{i=1}^2\sum_{j=1}^2 
\sqrt 2 g \frac{m_W}{m_t} \sin\beta (V_{i1}^*D_{t1j}^*-K_t V_{i2}^*D_{t2j}^*) 
\nonumber\\
&& (-\sqrt 2 X_{3k} U_{i1}^* +X_{2k} U_{i2}^* +\tan\theta_W X_{1k} U_{i2}^*) 
\nonumber\\
&& (\alpha_{tk}  D_{t1j}-\gamma_{tk}D_{t2j}) 
\frac{1}{16\pi^2} 
\frac{m_{\chi_k^0}m_{\chi_i^-}}{ m_{\tilde t_j}^2}
 H(\frac{m_{ \chi_i^-}^2}{m_{\tilde t_j}^2}, 
\frac{m_{\chi_k^0}^2}{m_{\tilde t_j}^2} )          
\eeqn

\noindent
{\bf Appendix D-Analysis of $\epsilon_{bb}$}\\

We  proceed now to compute the $\epsilon_{bb}$. It is given by
\beqn
\epsilon_{bb} = \sum_{i=1}^8 \epsilon_{bb}^{(i)}  
\eeqn
We exhibit the individual contributions below. 

\noindent
From Fig.\ref{epsfig3}(i) we find 
\beqn
\epsilon_{bb}^{(1)} &=& 
-\sum_{i=1}^2\sum_{j=1}^2 \frac{2\alpha_s}{3\pi} e^{-i\xi_3} D_{b1i}^*D_{b2j} 
[\mu^* D_{b1i}D_{b2j}^* 
+\frac{m_Zm_W}{m_b} \frac{\cos\beta}{\cos\theta_W} \\  
&& \{(-\frac{1}{2} +\frac{1}{3}\sin^2\theta_W)D_{b1i}D_{b1j}^*
- \frac{1}{3}\sin^2\theta_W D_{b2i}D_{b2j}^*\}\sin\beta ]
\frac{1}{|m_{\tilde g}|} H(\frac{m_{\tilde b_i}^2}{|m_{\tilde g}|^2}, 
\frac{m_{\tilde b_j}^2}{|m_{\tilde g}|^2}) \nonumber
\eeqn
\noindent
From Fig.\ref{epsfig3}(ii) we find 
the SUSY QCD  non-$\tan\beta$ enhanced contribution 
\beqn
\epsilon_{bb}^{(2)} &=& -\frac{1}{\tan\beta} 
\sum_{i=1}^2\sum_{j=1}^2 \frac{2\alpha_s}{3\pi} e^{-i\xi_3}
D_{b1i}^*D_{b2j} 
[-A_b D_{b1i}D_{b2j}^*  -m_b\{D_{b1j}^*D_{b1i} + D_{b2j}^*D_{b2i}\} 
\nonumber\\  && 
-\frac{m_Zm_W}{m_b} \frac{\cos\beta}{\cos\theta_W}
\{(-\frac{1}{2} +\frac{1}{3}\sin^2\theta_W) D_{b1i}  D_{b1j}^*
- \frac{1}{3}\sin^2\theta_W D_{b2i}D_{b2j}^*\}\cos\beta ]
\nonumber\\ &&
\frac{1}{|m_{\tilde g}|} H(\frac{m_{\tilde b_i}^2}{|m_{\tilde g}|^2}, 
\frac{m_{\tilde b_j}^2}{|m_{\tilde g}|^2})
\eeqn

\noindent
From Fig.\ref{epsfig3}(iii) we find 
\beqn
\epsilon_{bb}^{(3)} &=&
- \sum_{i=1}^2\sum_{j=1}^2 \sum_{k=1}^2 g^2 [
-\frac{m_t}{m_b} \cot\beta A_t^* D_{t2i}D_{t1j}^* 
-\frac{m_t^2}{m_b} \cot\beta \{D_{t1i}D_{t1j}^* + D_{t2i}D_{t2j}^*\} 
\nonumber\\
&& + \frac{m_Zm_W}{m_b} \frac{\cos\beta}{\cos\theta_W} 
\{(\frac{1}{2} -\frac{2}{3}\sin^2\theta_W) D_{t1i}  D_{t1j}^*
+ \frac{2}{3}\sin^2\theta_W  D_{t2i}D_{t2j}^*\}\sin\beta ]
\nonumber\\
&&(V_{k1}^*D_{t1i}^*-K_t V_{k2}^* D_{t2i}^* ) (K_b U_{k2}^* D_{t1j}) 
\frac{1}{16\pi^2}
\frac{1}{|m_{\tilde \chi_k^+}|} H(\frac{m_{\tilde t_i}^2}{|m_{\tilde\chi_K^+}|^2}, 
\frac{m_{\tilde t_j}^2}{|m_{\tilde \chi_k^+}|^2})
\eeqn

\noindent
From Fig.\ref{epsfig3}(iv),  which is non-$\tan\beta$ enhanced, we find 
\beqn
\epsilon_{bb}^{(4)} &=& - \frac{1}{\tan\beta}
\sum_{i=1}^2\sum_{j=1}^2 \sum_{k=1}^2 
g^2 [ \frac{m_t}{m_b} \cot\beta  \mu D_{t1j}^*D_{t2i} 
\nonumber\\  &&
-\frac{m_Zm_W}{m_b} \frac{\cos\beta}{\cos\theta_W} 
\{(\frac{1}{2} -\frac{2}{3}\sin^2\theta_W) D_{t1i}  D_{t1j}^*
+ \frac{2}{3}\sin^2\theta_W  D_{t2i}D_{t2j}^*\}\cos\beta]
\nonumber\\
&&(V_{k1}^*D_{t1i}^*-K_t V_{k2}^* D_{t2i}^* ) (K_b U_{k2}^* D_{t1j}) 
\frac{1}{16\pi^2}
\frac{1}{|m_{\tilde \chi_k^+}|} H(\frac{m_{\tilde t_i}^2}{|m_{\tilde\chi_k^+}|^2}, 
\frac{m_{\tilde t_j}^2}{|m_{\tilde \chi_k^+}|^2})
\eeqn

\noindent
From Fig.\ref{epsfig3}(v) we find 
\beqn
\epsilon_{bb}^{(5)} &=&
\sum_{i=1}^2\sum_{j=1}^2 \sum_{k=1}^4 
2 [\mu^*  D_{b1i}D_{b2j}^* + 
\frac{m_Zm_W}{m_b} \frac{\cos\beta}{\cos\theta_W} 
\nonumber\\  
&& \{(-\frac{1}{2} +\frac{1}{3}\sin^2\theta_W) D_{b1i}  D_{b1j}^*
- \frac{1}{3}\sin^2\theta_W  D_{b2i}D_{b2j}^*\}\sin\beta]
\nonumber\\
&& (\alpha_{bk}D_{b1j}-\gamma_{bk}D_{b2j})
(\beta_{bk}^* D_{b1i}^*+\alpha_{bk}D_{b2i}^*)
\frac{1}{16\pi^2}
\frac{1}{|m_{\tilde \chi_k^0}|} H(\frac{m_{\tilde b_i}^2}{|m_{\tilde\chi_k^0}|^2}, 
\frac{m_{\tilde b_j}^2}{|m_{\tilde \chi_k^0}|^2})
\eeqn
\noindent
From Fig.\ref{epsfig3}(vi), which is non-$\tan\beta$ enhanced, we find 
\beqn
\epsilon_{bb}^{(6)} &=& \frac{1}{\tan\beta} 
\sum_{i=1}^2\sum_{j=1}^2 \sum_{k=1}^4 
2 [-A_b D_{b1i}D_{b2j}^*-m_b  \{ D_{b1i}D_{b1j}^* +  D_{b2i}D_{b2j}^* \}
\nonumber\\  
&&-\frac{m_Zm_W}{m_b} \frac{\cos\beta}{\cos\theta_W} 
\{(-\frac{1}{2} +\frac{1}{3}\sin^2\theta_W) D_{b1i}  D_{b1j}^*
- \frac{1}{3}\sin^2\theta_W  D_{b2i}D_{b2j}^*\}\cos\beta] 
\nonumber\\
&&(\alpha_{bk}D_{b1j}-\gamma_{bk}D_{b2j})
(\beta_{bk}^* D_{b1i}^*+\alpha_{bk}D_{b2i}^*)
\frac{1}{16\pi^2}
\frac{1}{|m_{\tilde \chi_k^0}|} H(\frac{m_{\tilde b_i}^2}{|m_{\tilde\chi_k^0}|^2}, 
\frac{m_{\tilde b_j}^2}{|m_{\tilde \chi_k^0}|^2})
\eeqn

\noindent
From Fig.\ref{epsfig4}(i) we find 
\beqn
\epsilon_{bb}^{(7)} &=&
\sum_{i=1}^2\sum_{j=1}^2 \sum_{k=1}^2 
2g^2 \frac{m_W}{m_b}\cot\beta [\frac{m_{\tilde \chi_i^+}}{2m_W}\delta_{ij} -Q_{ij}^*\cos\beta -R_{ij}^*] 
\\  
&& (V_{i1}^*D_{t1k}^*-K_t V_{i2}^*D_{t2k}^*) (K_b U_{j2}^*D_{t1k}) 
\frac{1}{16\pi^2}
\frac{|m_{\tilde \chi_i^+}||m_{\tilde \chi_j^+}|}{m_{\tilde t_k}^2} 
 H(   \frac{|m_{\tilde\chi_i^+}|^2} {m_{\tilde t_k}^2}, 
\frac{|m_{\tilde \chi_j^+}|^2} {m_{\tilde t_k}^2} )   \nonumber
\eeqn
\noindent
From Fig.\ref{epsfig4}(ii) we find 
\beqn
\epsilon_{bb}^{(8)} &=&
-\sum_{i=1}^4\sum_{j=1}^4 \sum_{k=1}^2 
4 \frac{m_W}{m_b}\cot\beta [\frac{m_{\tilde
 \chi_i^0}}{2m_W}\delta_{ij} -Q_{ij}^{''*}\cos\beta - R_{ij}^{''*}] 
(\alpha_{bj}D_{b1k} - \gamma_{bj}D_{b2k})  
\nonumber\\  && 
(\beta_{bi}^*D_{b1k}^* + \alpha_{bi} D_{b2k}^* ) 
\frac{1}{16\pi^2}
\frac{|m_{\tilde \chi_i^0}||m_{\tilde \chi_j^0}|}{m_{\tilde b_k}^2} 
 H(   \frac{|m_{\tilde\chi_i^0}|^2} {m_{\tilde b_k}^2}, 
\frac{|m_{\tilde \chi_j^0}|^2} {m_{\tilde b_k}^2} )     
\eeqn
In the above $Q, R, Q"$ and $R"$ are defined by 
\beqn
Q_{ij} &=& \sqrt{\frac{1}{2}} U_{i2}V_{j1}\nonumber\\
R_{ij} &=& \frac{1}{2M_W} [\tilde m_2^* U_{i1} V_{j1} + \mu^* U_{i2} V_{j2}] 
\eeqn
and by
\beqn
gQ^{''}_{ij} &=& \frac{1}{2} [ X_{3i}^* (gX_{2j}^* -g' X_{1j}^*) +
(i\leftarrow \rightarrow j) ]\nonumber\\
R_{ij}^{''}  &=& \frac{1}{2M_W} [ \tilde m_1^* X_{1i}^*X_{1j}^* 
+ \tilde m_2^* X_{2i}^*X_{2j}^*
-\mu^* (X_{3i}^*X_{4j}^* + X_{4i}^*X_{3j}^*) ] 
\eeqn

\begin{figure}
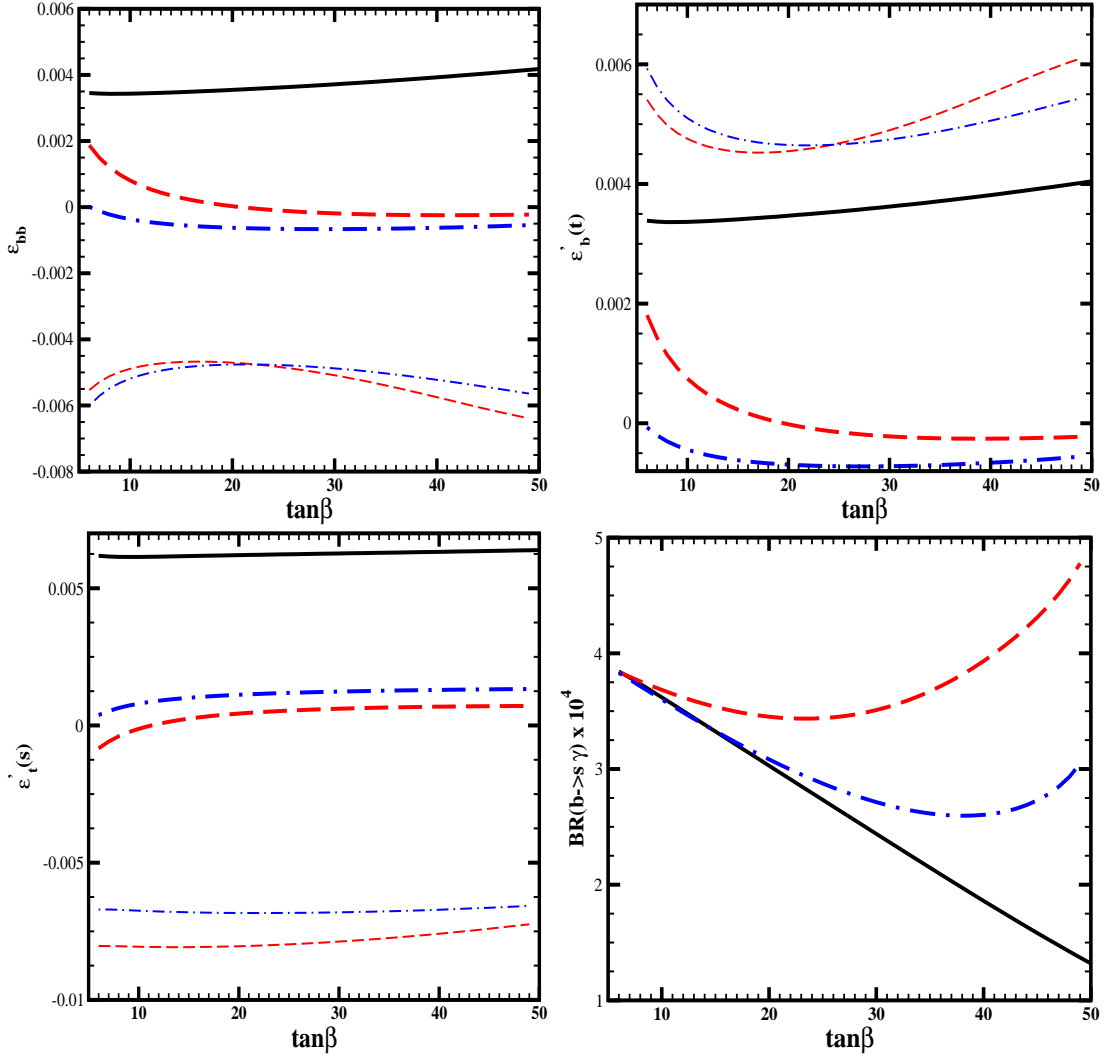

\begin{center}
\includegraphics[width=7.2cm,height=7cm]{graphs/ebb_tb.eps}
\includegraphics[width=7.2cm,height=7cm]{graphs/ebt_tb.eps}
\includegraphics[width=7.2cm,height=7cm]{graphs/ets_tb.eps}
\includegraphics[width=7.2cm,height=7cm]{graphs/bsg_tb0.eps}
\end{center}
\caption{Variation of the $\epsilon$'s and BR($b\rightarrow s \gamma$) 
with $\tan\beta$ for the parameters $m_0= 400$~GeV, $m_{1/2}= 200$~GeV, 
$A_0=700$~GeV and all the phases set to zero (solid lines) and with the 
phases (in radiants) set to $\xi_1=0$, $\xi_2=0.8$, $\xi_3=0.9$, 
$\theta_\mu=0.6$, $\alpha_0=\pi$ (dash lines) or  
$\xi_1=1$, $\xi_2=0$, $\xi_3=0.3$, 
$\theta_\mu=1.2$, $\alpha_0=\pi/2$ (dot-dash lines). The thick lines represent 
real values while the thin lines correspond to imaginary parts.}
\label{bsgphases}
\end{figure}

\begin{figure}
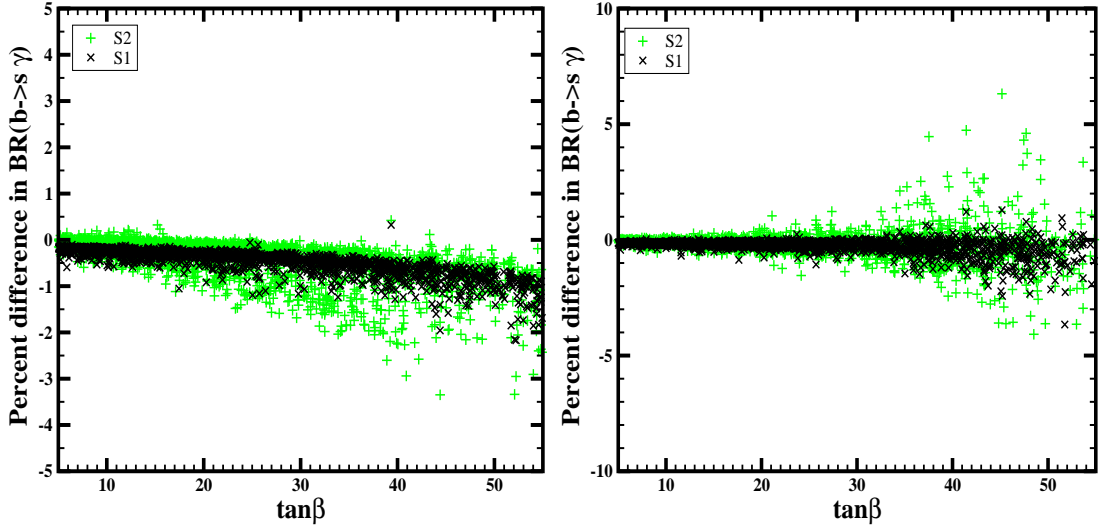

\includegraphics[width=7.2cm,height=7cm]{graphs/real.eps}
\includegraphics[width=7.2cm,height=7cm]{graphs/phases.eps}
\caption{The percentage difference  between the approximate formulae 
of $S_1$ and $S_2$ and  our full calculation in the SUGRA 
scenario. The left graph is the  case with no phases, 
and the right graph is the case with phases. Each group contains about 1800 models where each point in the parameter space defines a model.}
\label{bsgsugra}
\end{figure}
\begin{figure}
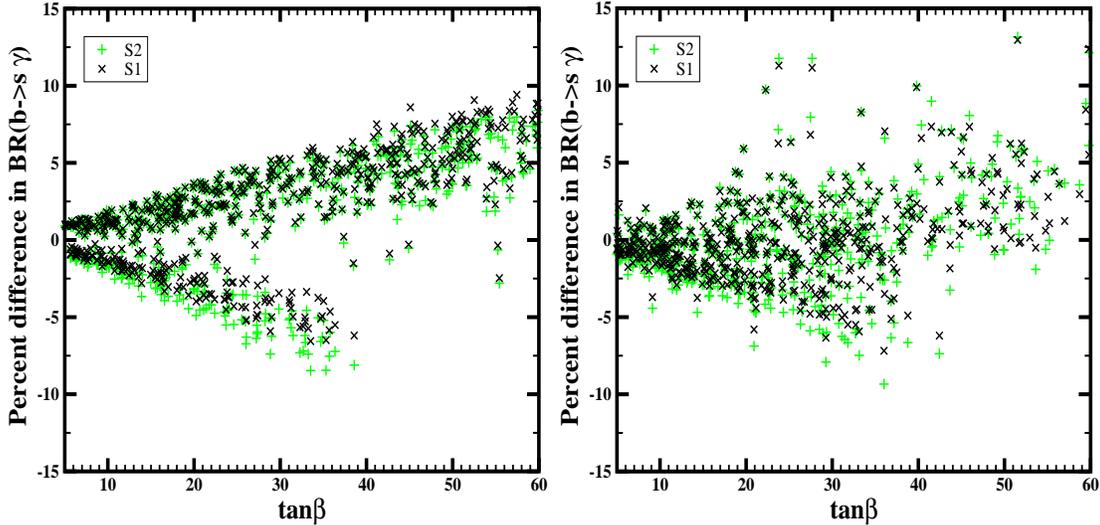

\begin{center}
\includegraphics[width=7.2cm,height=7cm]{graphs/realsugra_wrong.eps}
\includegraphics[width=7.2cm,height=7cm]{graphs/compsugra_wrong.eps}
\end{center}
\caption{The percentage difference  between the non-corrected 
approximate formulae as they appear in Ref.\cite{Demir:2001yz} and 
Ref.\cite{Belanger:2004yn}  and our full calculation in the real 
mSUGRA case which has no phases (left) and in the complex SUGRA case which has phases (right).}
\label{bsgsugrawrong}
\end{figure}

\begin{figure}
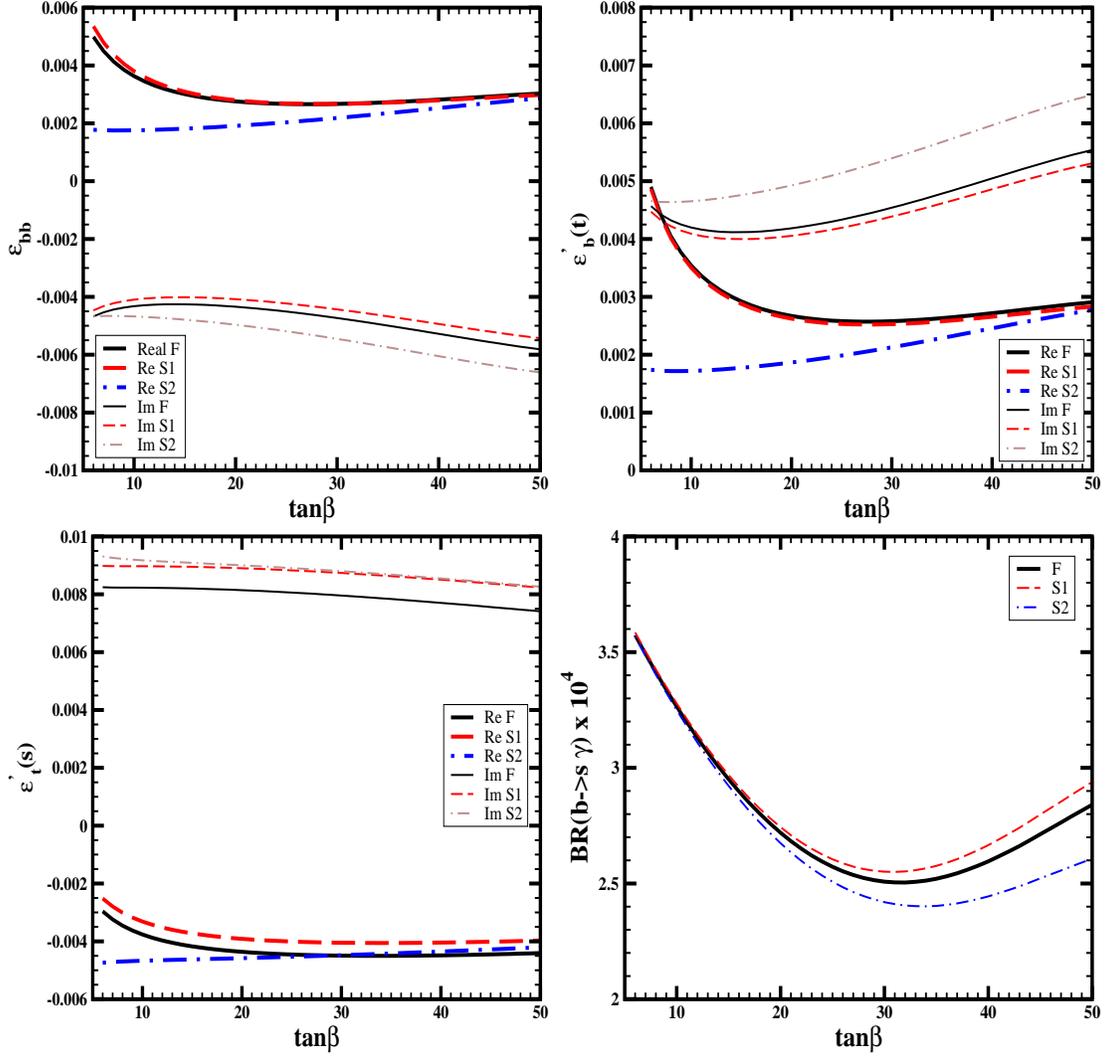

\begin{center}
\includegraphics[width=7.2cm,height=7cm]{graphs/ebb.eps}
\includegraphics[width=7.2cm,height=7cm]{graphs/epbt.eps}
\includegraphics[width=7.2cm,height=7cm]{graphs/epts.eps}
\includegraphics[width=7.2cm,height=7cm]{graphs/bsgtb.eps}
\end{center}
\caption{Variation of the $\epsilon$'s with $\tan\beta$ for the
parameters of Eq.~(\ref{point0}) and the corresponding prediction of 
BR($b \rightarrow s \gamma$) using $S_1$, $S_2$ and 
the complete calculation ($F$).}
\label{pointsugra}
\end{figure}


\begin{figure}
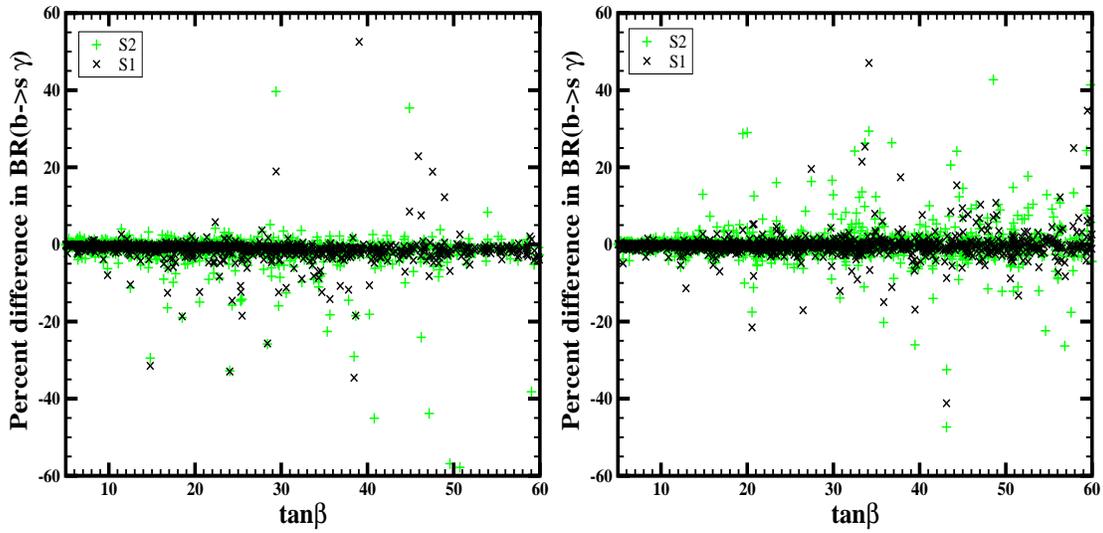

\begin{center}
\includegraphics[width=7.2cm,height=7cm]{graphs/realmssm.eps}
\includegraphics[width=7.2cm,height=7cm]{graphs/compmssm.eps}
\end{center}
\caption{The percentage difference  between the approximate formulae 
of $S_1$ and  $S_2$ and our full calculation in the MSSM 
scenario. The data sets contain about 1000 models. 
We only plot points that are experimentally acceptable.
Left graph is the case with no phases, and right graph is the case with phases.}
\label{bsgmssm}
\end{figure}

\begin{figure}
\begin{center}
\includegraphics[width=7.2cm,height=7cm]{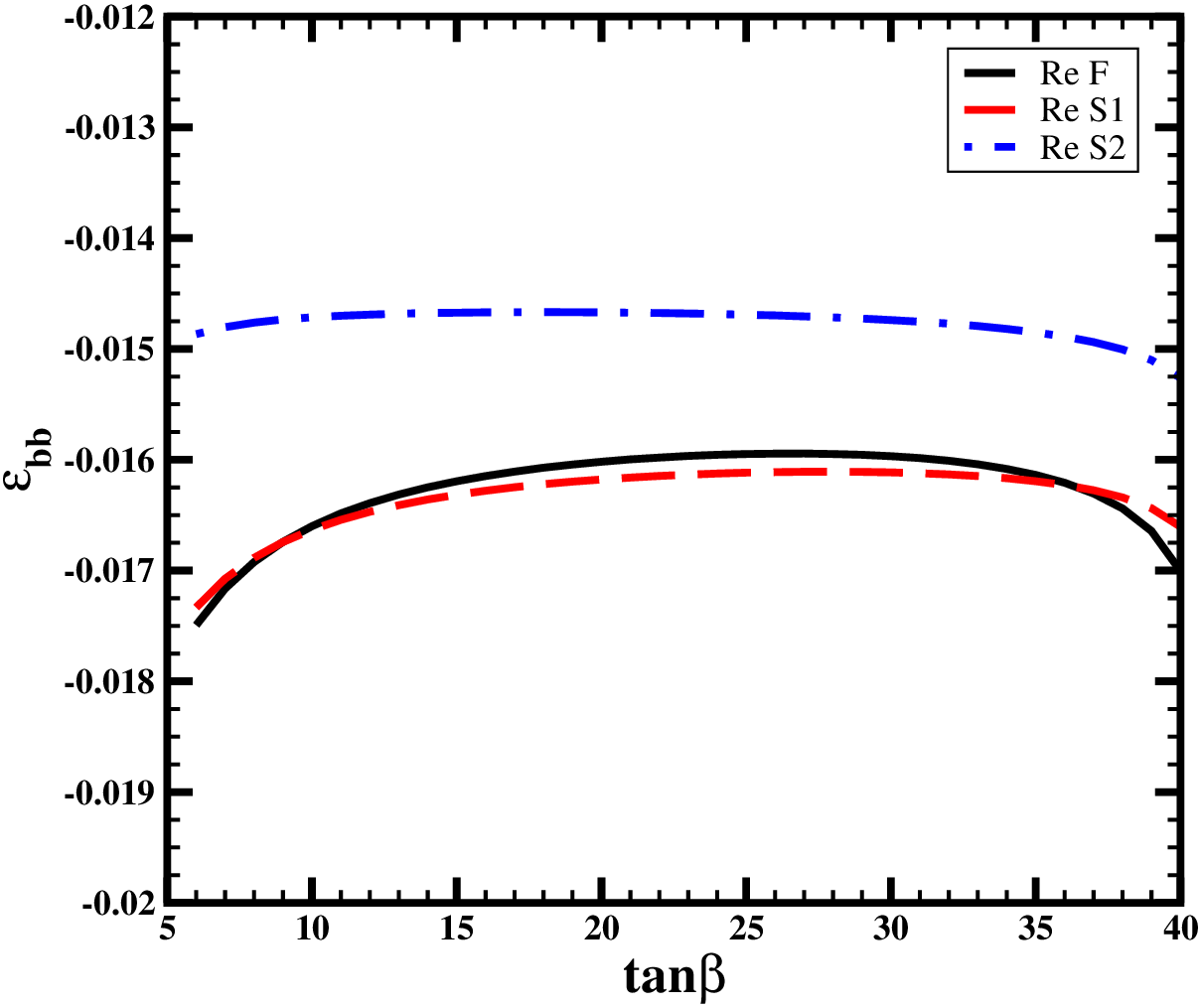}
\includegraphics[width=7.2cm,height=7cm]{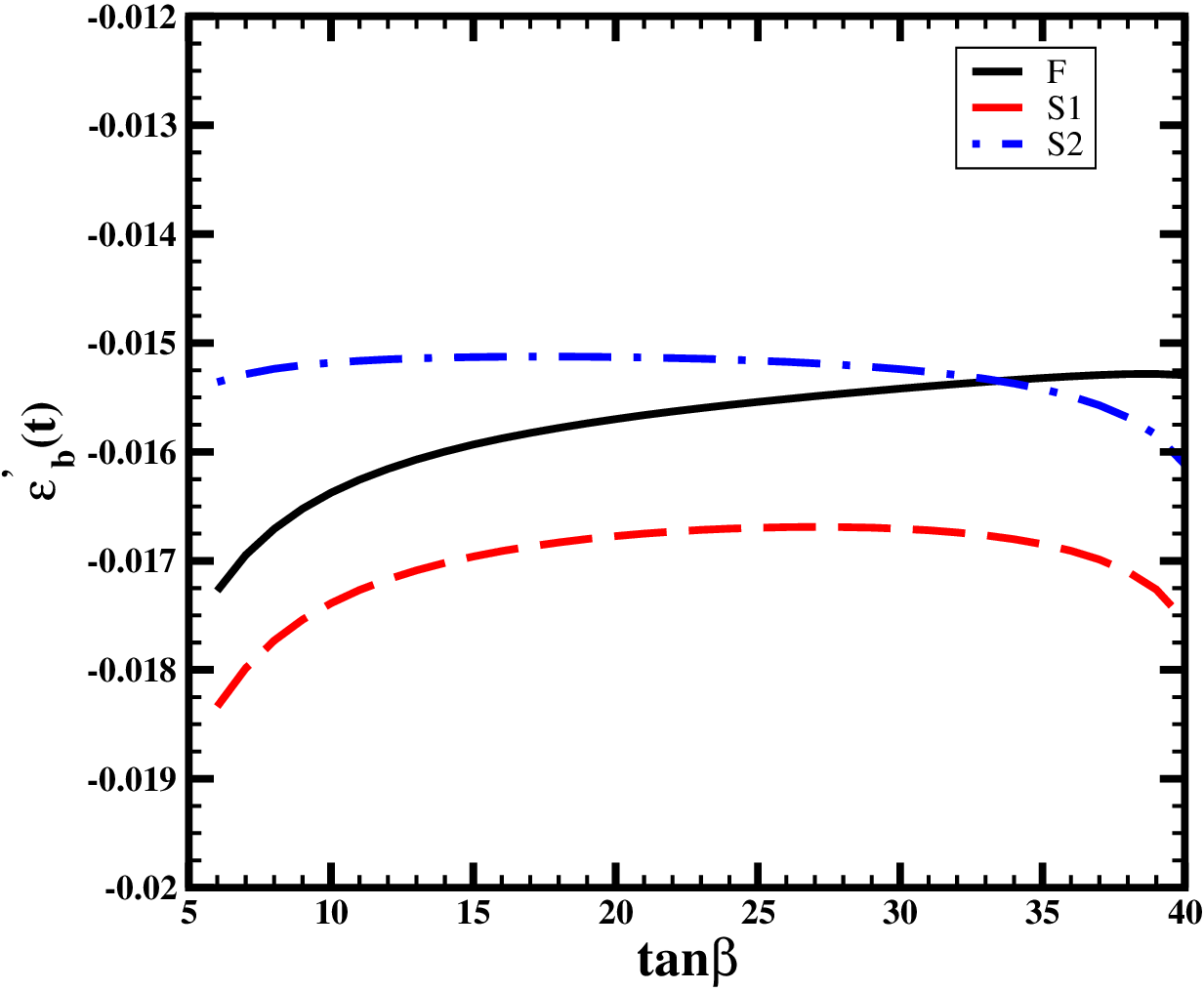}
\includegraphics[width=7.2cm,height=7cm]{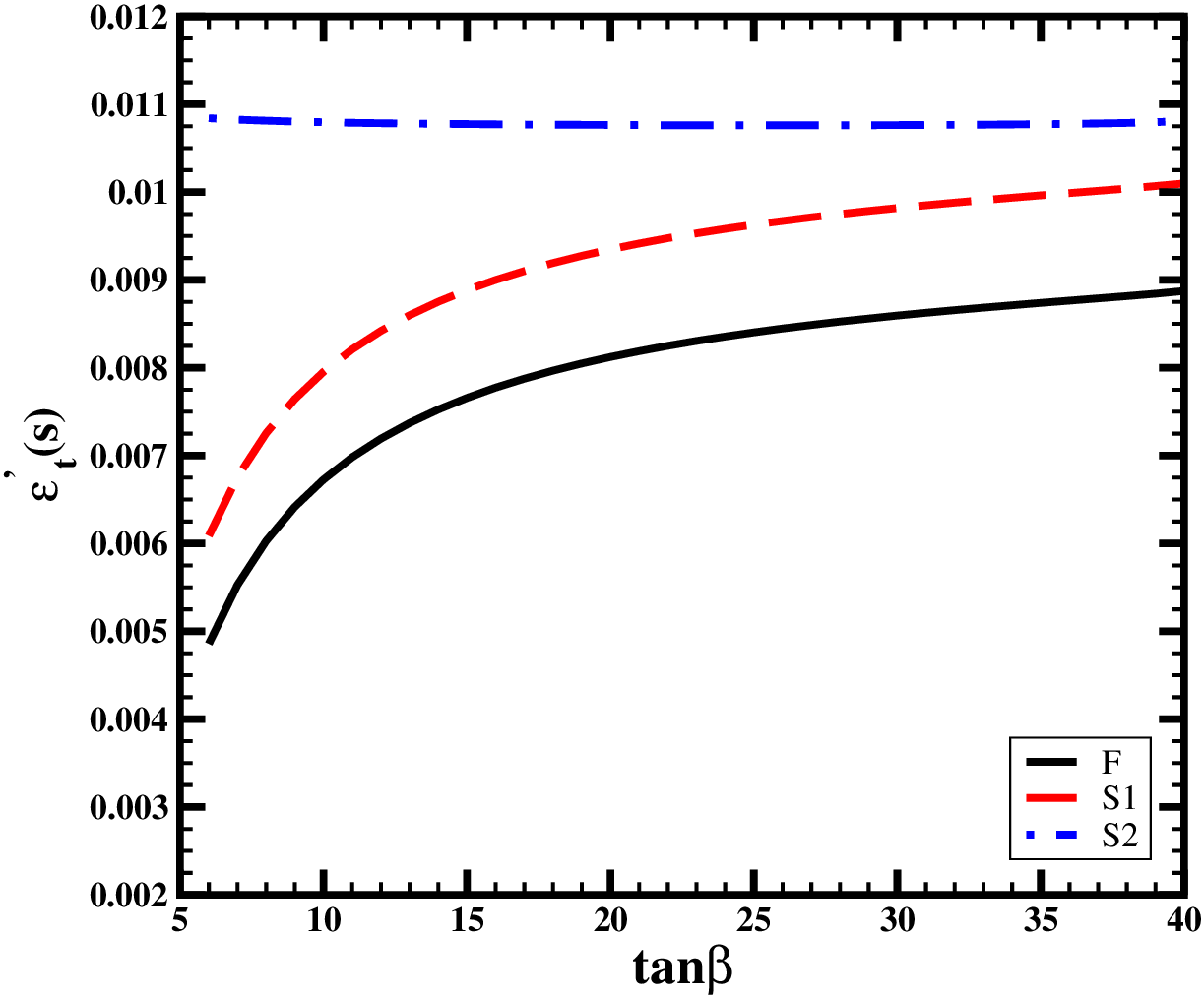}
\includegraphics[width=7.2cm,height=7cm]{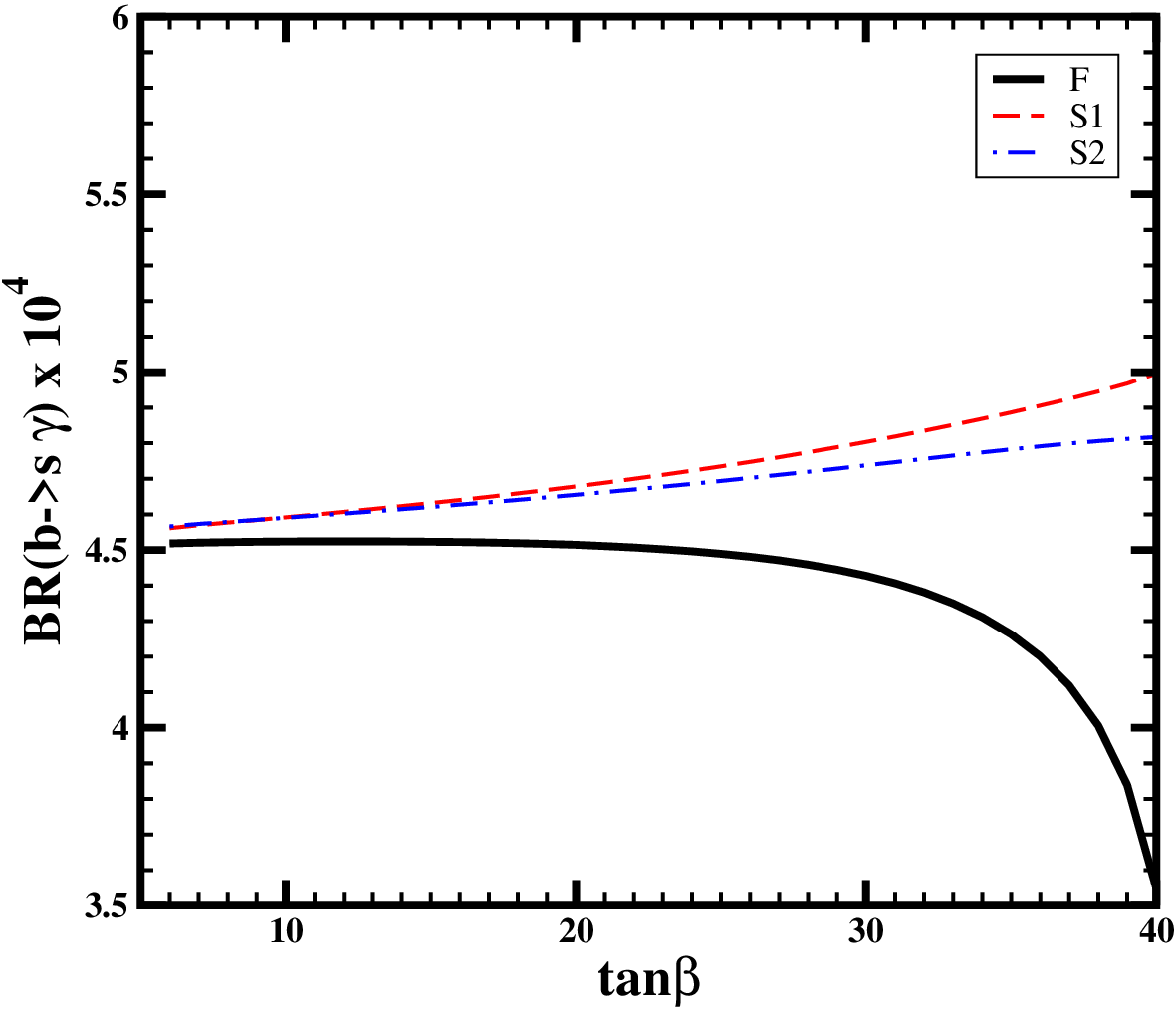}
\end{center}
\caption{The values of the $\epsilon$'s and the rate for 
$b \rightarrow s \gamma$ for the three different methods 
at point point (i).}
\label{point1plot}
\end{figure}

\begin{figure}
\begin{center}
\includegraphics[width=7.2cm,height=7cm]{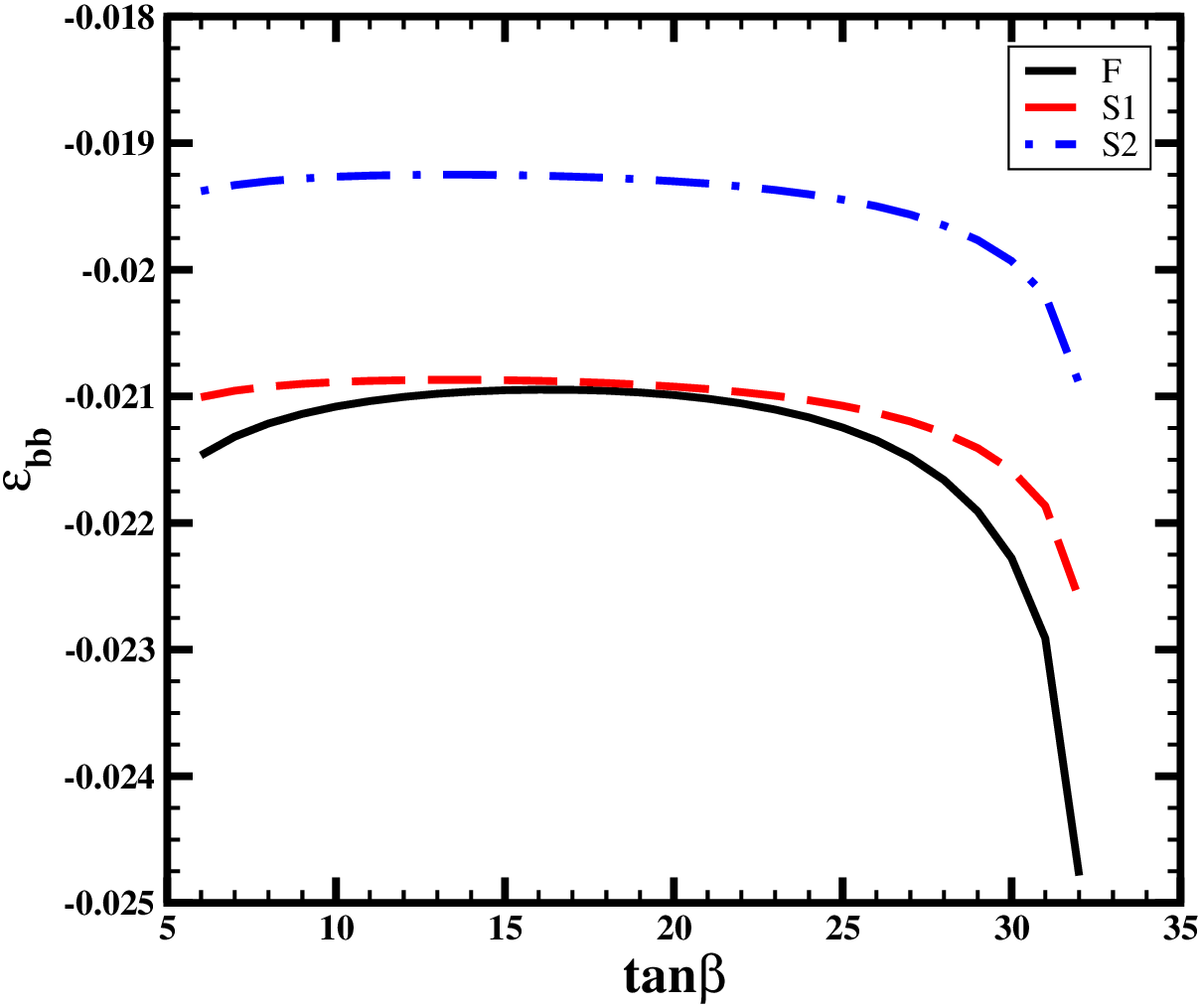}
\includegraphics[width=7.2cm,height=7cm]{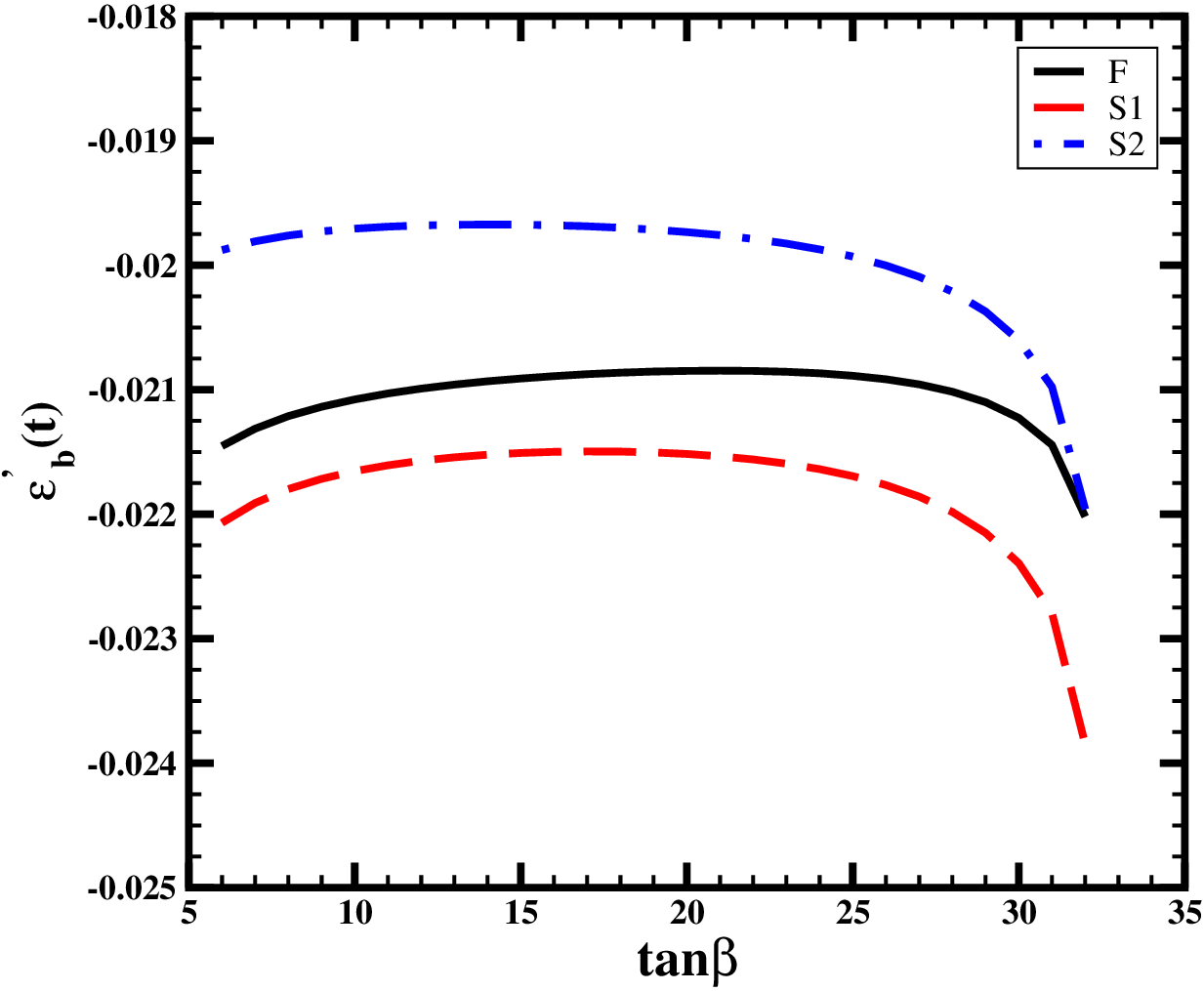}
\includegraphics[width=7.2cm,height=7cm]{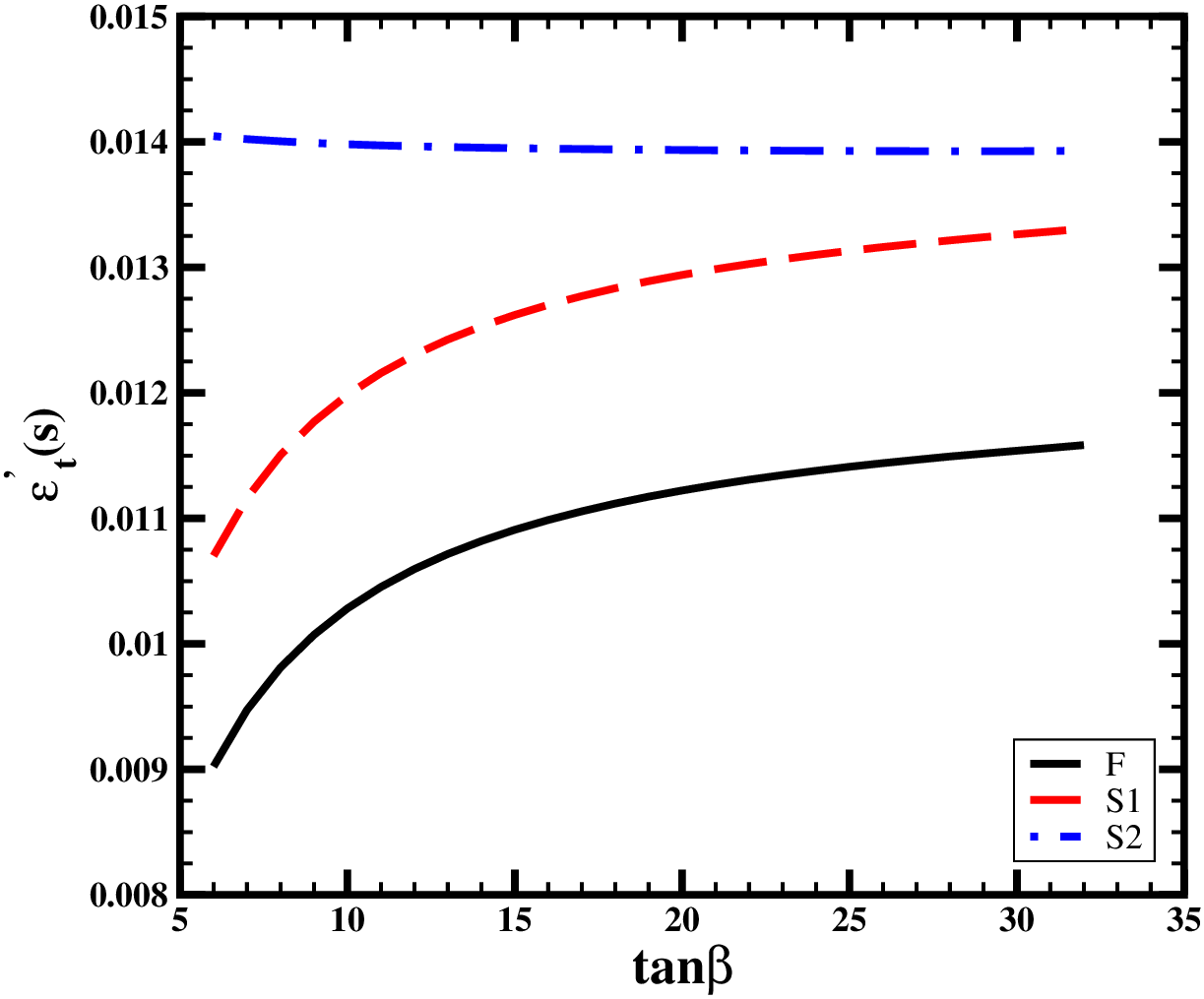}
\includegraphics[width=7.2cm,height=7cm]{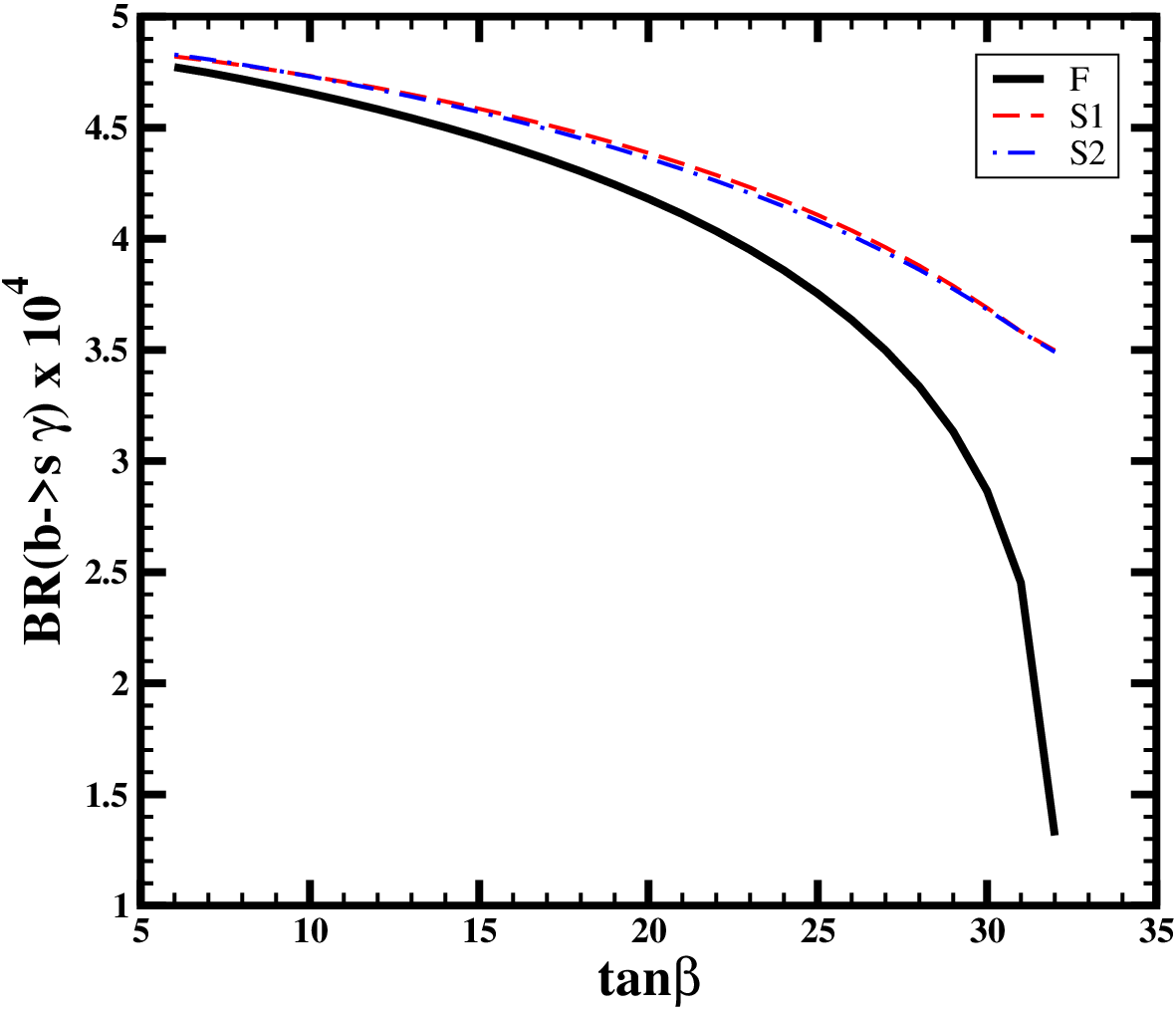}
\end{center}
\caption{The values of the $\epsilon$'s and the rate for 
$b \rightarrow s \gamma$ for the three different methods 
at point point (ii).}
\label{point2plot}
\end{figure}

\end{document}